\documentclass[aps,prb,twocolumn,showpacs,floatfix,superscriptaddress]{revtex4}
\usepackage{dcolumn}
\usepackage{bm}
\usepackage{amsmath,amssymb,graphicx}

\begin{document}
\preprint{Submitted to Phys. Rev. B}


\title{Resonant Coherent Phonon Spectroscopy of Single-Walled Carbon Nanotubes}

\author{G. D. Sanders}
\author{C. J. Stanton}
\affiliation{Department of Physics, University of Florida, Box 118440,
Gainesville, Florida 32611-8440}

\author{J.-H.~Kim}
\affiliation{Department of Physics, Chungnam National University, Daejeon, 305-764, Republic of Korea}

\author{K.-J.~Yee}
\affiliation{Department of Physics, Chungnam National University, Daejeon, 305-764, Republic of Korea}

\author{Y.-S.~Lim}
\affiliation{Department of Applied Physics, Konkuk University, Chungju, Chungbuk, 380-701, Republic of Korea}

\author{E.~H.~H\'{a}roz}
\author{L.~G.~Booshehri}
\affiliation{Department of Electrical and Computer Engineering, Rice University, Houston, Texas 77005, USA}

\author{J.~Kono}
\affiliation{Department of Electrical and Computer Engineering, Rice University, Houston, Texas 77005, USA}

\author{R. Saito}
\affiliation{Department of Physics, Tohoku University,
Sendai 980-8578, Japan}

\date{\today}


\begin{abstract}
Using femtosecond pump-probe spectroscopy with pulse shaping techniques, one can
generate and detect coherent phonons in chirality-specific semiconducting single-walled
carbon nanotubes.  The signals are resonantly enhanced when the pump photon energy
coincides with an interband exciton resonance, and analysis of such data provides a
wealth of information on the chirality-dependence of light absorption, phonon generation,
and phonon-induced band structure modulations.  To explain our experimental results,
we have developed a microscopic theory for the generation and
detection of coherent phonons in single-walled carbon nanotubes using a
tight-binding model for the electronic states and a valence force field model for
the phonons.  We find that the coherent phonon amplitudes satisfy a driven oscillator
equation with the driving term depending on photoexcited carrier density.
We compared our theoretical results with experimental results on mod~2 nanotubes and
found that our model provides satisfactory overall trends in the relative strengths
of the coherent phonon signal both within and between different mod~2 families.
We also find that the coherent phonon intensities are considerably weaker in
mod~1 nanotubes in comparison with mod~2 nanotubes, which is also in excellent
agreement with experiment.
\end{abstract}

\pacs{78.67.Ch, 63.22.+m, 73.22.-f, 78.67.-n}


\maketitle

\section{Introduction}
\label{Introduction section}

Single-walled carbon nanotubes (SWNT) can be viewed as rolled up sheets of graphene,
having a one-dimensional band structure with unique electronic, mechanical, and optical
properties.  Their electronic properties vary significantly, depending on their chirality
indices ($n$,$m$), and can be either metallic or semiconducting.
\cite{Saito2003,Dresselhaus2001,Harris1999,Charlier07.677,Terrones03.419}
Although there are currently world-wide efforts to achieve single-chirality samples,
a standard for fabrication of such samples has yet to be established. Resonant Raman
spectroscopy (RRS) or photoluminescence excitation spectroscopy (PLE) is
usually used to study chirality-dependent electronic and vibrational properties.
However, carbon nanotube samples typically contain ensembles of nanotubes with different
chiralities, and the unknown relative abundances of different-chirality tubes in such
samples often makes it challenging to extract reliable parameters on chirality-dependent
properties from experimental results.

Resonant Raman spectroscopy can be used to study chirality-dependent electron-phonon
coupling in nanotubes and can be used to uniquely determine the chirality of individual
tubes.\cite{Jorio01.1118,Jorio03.139,Doorn04.1147,Jiang07.035405,Jiang07.035407}
Raman spectroscopy is a sensitive
probe of ground-state vibrations but is less suitable for studying excited state vibrational
properties.  Recently, excited state lattice vibrations in carbon nanotubes have been studied
with coherent phonon (CP) spectroscopy.\cite{Gambetta06.515,Lim06.2696,Lim07.306,KatoetAl08NL}
In CP spectroscopy, coherent phonon oscillations are excited by pumping with an ultrafast
pump pulse and are detected by measuring changes in the differential transmission using a
delayed probe pulse.  The CP intensity is then obtained by taking the temporal power spectrum
of the differential transmission.  The peaks in the power spectrum correspond to coherent phonon
frequencies.  Coherent phonon spectroscopy allows direct measurement of excited state phonon
dynamics in the time domain including phase information and dephasing times.

We have developed a technique that allows us to study chirality dependent properties
of nanotubes in an ensemble.\cite{Kim09.037402} This is described in
Section \ref{Experiment section}. By shaping the pump pulse, we incorporate quantum
control techniques in CP spectroscopy.  Using pre-designed trains of femtosecond optical
pulses, we have selectively excited and probed coherent lattice vibrations of the radial
breathing mode (RBM) of specific chirality single-walled carbon nanotubes.  We are
able to gain information on light absorption, coherent phonon generation, and
coherent phonon-induced band structure modulations.  We find that coherent RBM
phonons can be selectively excited by using a train of pump pulses whose repetition
rate is in resonance with the desired phonon frequency.  By exciting only those phonon
modes with a specific frequency, we can selectively study nanotubes with the same
chirality in an ensemble of tubes.

In order to explain our experimental results, we develop in
Section \ref{Theory section} a microscopic theory for the generation and
detection of coherent phonon lattice vibrations in carbon nanotubes by ultrafast
laser pulses. We use a third nearest neighbor extended tight-binding (ETB) model to
describe the electronic states over the entire Brillouin zone while the phonons
are treated in a valence force field model. In treating the electrons and phonons,
we exploit the screw symmetry of the nanotube to drastically simplify the problem.
Equations of motion for each CP vibrational mode are obtained, using a microscopic
description of the electron-phonon interaction based on direct evaluation of the
three-center electron-phonon matrix elements using {\em ab initio} wavefunctions
and screened atomic potentials. For each CP active mode we find that the CP amplitudes
satisfy a driven oscillator equation with a coherent phonon driving function that
depends on photoexcited hot carrier distributions.  An ultrafast laser pulse generates
electron-hole pairs and the driving function rises sharply in a step-like fashion.
If the pulse duration is shorter than the phonon oscillation period, the rapid
initial jump in the coherent phonon driving function gives rise to oscillating
coherent phonon amplitudes.

Carbon nanotubes with the same values of $2 n + m$ are said to belong to the
same family (with index $2 n + m$). Carbon nanotubes in a given family
are metallic if $\mod(n-m,3)=0$ and semiconducting otherwise. The semiconducting
tubes are classified as either mod~1 or mod~2  depending on whether
the value of $\mod(n-m,3)$ is 1 or 2. In CP spectroscopy, we find that a strong
signal is obtained when we pump at the allowed nanotube $E_{ii}$ optical transitions.
We found experimentally that, for the RBM modes, the CP intensity within a mod~2
family tends to decrease with chiral angle and the decrease in CP intensity with
chiral angle is found to be much more pronounced for the $E_{11}$ feature.  We also
found that CP intensities are considerably weaker in mod~1 families in comparison
with mod~2 families. In general, the $E_{22}$ CP intensities in mod~2 families are
stronger than the $E_{11}$ features while the opposite is true in mod~1 tubes.
For RBM modes in mod~1 tubes, the $E_{11}$
CP intensities tend to decrease with increasing chiral angle within a given family.
As the family index (2$n$+$m$) increases, the $E_{11}$ CP intensity in mod~1 tubes
decreases.  Finally, we compared our theoretical results with experimental CP
spectra in mod~2 nanotubes and found that our theoretical model correctly predicts
the experimentally observed overall trends in the relative strengths of the CP signal
both within and between mod~2 families.  We found discrepancies between our
theoretical predictions with regard to the peak positions and lineshapes.
These discrepancies can be qualitatively attributed to Coulomb interactions which
have not yet been included in the calculations.

We do not consider the Coulomb interaction and excitonic effects in our theoretical
model for reasons of simplicity and tractability. It has been pointed out that
excitonic effects are important for understanding the optical properties of small
diameter carbon nanotubes.\cite{Spataru92.077402} In a number of Raman scattering
theories the Coulomb interaction is neglected, but nevertheless the computed RBM Raman
spectra can explain many experimental measurements.\cite{Jiang05.205420,Machon05.035416,Popov04.17}
It is worth dwelling a little on the reason for this. Jiang \textit{et al}.\cite{Jiang07.035405}
recently undertook a study of the exciton-photon and exciton-phonon matrix elements in
single-walled carbon nanotubes using a tight-binding model. These authors found that
for the RBM and G-band modes, the phonon matrix elements in the exciton and free
particle pictures are nearly the same. However, values for the exciton-photon
matrix elements are on the order of 100 times greater than the electron-photon
matrix elements computed in the free particle picture. Thus, when we discuss
the photoexcitation of carriers, the actual photoexcited carrier densities will
be different from what we predict. On the other hand when we discuss the
dependence of the coherent phonon amplitudes on the $E_{ii}$ transition energies
or the tube chiraility, the present discussion has a physical meaning.
Thus we expect reasonable agreement between experiment and theory for relative
coherent phonon amplitudes and the relative strengths of the computed CP
spectra. However, the peak positions and line shapes of the CP spectra will be
altered by the neglect of excitonic effects.

Apart from being useful in resonant CP spectroscopy of chirality specific
nanotubes, we note that
laser induced coherent phonons in carbon nanotubes may also have important
practical applications in the fabrication of carbon nanotube
electronic devices. In recent years laser induced coherent phonon generation has
been theoretically studied using molecular dynamics techniques.
\cite{Garcia04.855,Dumitrica04.117401,Dumitrica06.193406,Valencia06.075409,Romero05.1361}
Garcia \textit{et al.}\cite{Garcia04.855} have simulated laser induced coherent
phonons in a mod 1 zigzag (10,0) capped carbon nanotube using a formalism that combines
a nonadiabatic molecular dynamics method and a density matrix approach to
describe the dynamics of the carbon ions and valence electrons.
Dumitric$\breve{a}$ \textit{et al.}\cite{Dumitrica04.117401,Dumitrica06.193406}
have theoretically studied the possibility of achieving selective cap opening
in (10,0), (5,5) and (8,4) capped carbon nanotubes driven by laser induced
coherent phonons using nonadiabatic molecular dynamics simulations based
on a microscopic electronic model.

It is well known that self-assembled carbon nanotubes suffer from structural
imperfections that modify their electronic, optical, and mechanical properties.
Such defects pose a problem for the fabrication of nanotube based electronic devices.
\cite{Romero05.1361} A common type of defect in carbon nanotubes is the (5-7) pair
defect introduced by applying a Stone-Wales transformation to the nanotube
structure.\cite{Valencia06.075409} In the Stone-Wales transformation, four
hexagons are replaced by two pentagons and two heptagons.\cite{Stone86.501}
The possibility of eliminating such defects using laser generated coherent phonons
has been studied theoretically in armchair and zigzag nanotubes using nonabiabatic
molecular dynamics simulations. Romero \textit{et al.}\cite{Romero05.1361} have
studied the response of armchair nanotubes with (5-7) pair defects to ultrafast
laser pulses and found that when the fraction of photoexcited electrons exceeds
a critical threshold ( around 7\% ) the resulting coherent phonon oscillations
cause the nanotube to undergo an inverse Stone-Wales structural transition that
heals the defect. More recently Valencia \textit{et al.}\cite{Valencia06.075409}
have extended these studies to zigzag nanotubes and found similar results.

Recently, Jeschke \textit{et al.}\cite{Jeschke07.125412} have theoretically
studied the structural response of nanotubes of different chiralities to
femtosecond laser excitation using molecular dynamics simulations. They found that
carbon nanotubes may transform into more stable structures under the appropriate
conditions. Such investigations may be important for technological applications.
For example, nanotubes excited by lasers above a certain threshold may tear open
and interact with other tubes leading to the creation of new structures.

\section{Experiment}
\label{Experiment section}
%
\begin{figure} [tbp]
\begin{center}
\includegraphics[scale=.7]{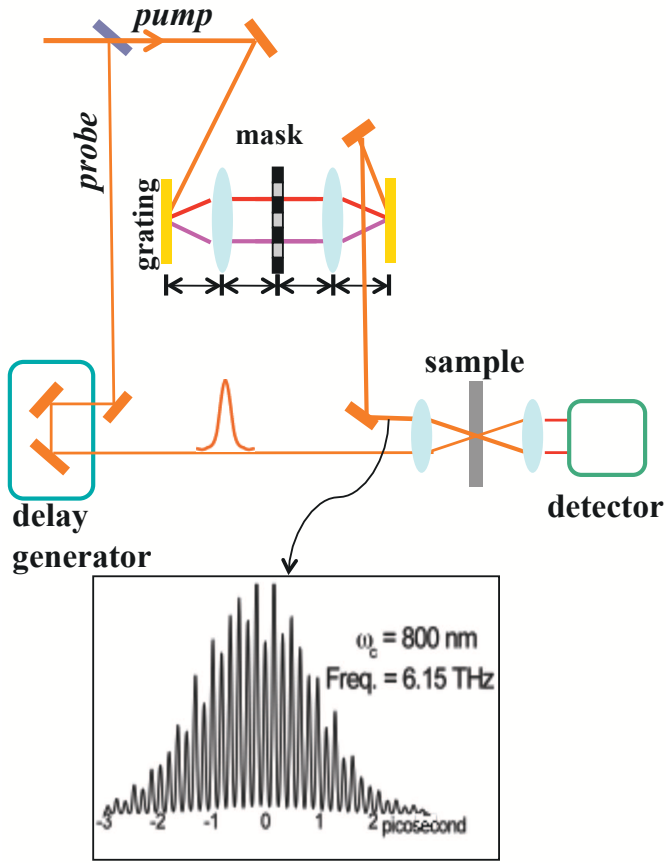}
\caption{(color online) Experimental setup for chirality-selective excitation
of coherent lattice vibrations in single-walled carbon nanotubes through
ultrafast pulse shaping.  The inset shows an example of multiple pulse trains
with a repetition rate of 6.15~THz and a central wavelength of 800~nm (1.55~eV).
}
\label{setup}
\end{center}
\end{figure}
%
In this section, we demonstrate the implementation of ultrafast pulse-shaping
to excite the coherent radial breathing modes of specific chiralities, providing
a definitive ability to study single chirality nanotubes from an ensemble sample.
Our method exploits selective excitation to not only extract chiral-dependent
band gap modulations, but also utilizes information from the probe energy
dependence of the phase and the amplitude of the coherent phonon oscillations to
reconstruct excitation profiles for the $E_{22}$ transitions. In particular, our
observation of probe-energy-dependent phase reversal provides direct, time-domain
evidence that for coherent radial breathing modes the band gap oscillates in
response to the nanotube diameter oscillations.

The sample used in this study was a micelle-suspended SWNT solution, where the
single-walled carbon nanotubes (HiPco batch HPR 104) were suspended as individuals
with sodium cholate.  The optical setup was that of standard degenerate pump-probe
spectroscopy, but chirality selectivity of RBM oscillations was achieved by using
multiple pulse trains, with a pulse-to-pulse interval corresponding to the period
of a specific RBM mode.~\cite{Kim09.037402}  Among different species of nanotubes,
those having RBM frequencies that are matched to the repetition rate of multiple
pulse trains will generate large amplitude coherent oscillations with increasing
oscillatory response to each pulse, while others will have diminished coherent
responses.\cite{Gambetta06.515,Lim06.2696,Lim07.306} The tailoring of multiple
pulse trains from femtosecond pulses was achieved using the pulse-shaping
technique developed by Weiner and Leaird.\cite{WeinerLeaird90OL}  As depicted
in Fig.~\ref{setup}, pulse trains are incident on an ensemble of nanotubes as
a pump beam, whereas coherent RBM oscillations are monitored by an unshaped,
Gaussian probe beam.
\begin{figure} [tbp]
\includegraphics[scale=1.65]{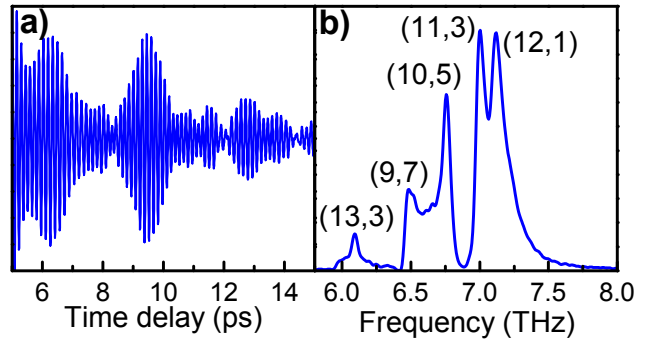}
\caption{(color online) Generation and detection of coherent phonons of the
radial breathing mode in single-walled carbon nanotubes.  (a) Time-domain
transmission modulations due to coherent RBM vibrations in ensemble single-walled
carbon nanotube solution that were generated by standard pump-probe spectroscopy
without pulse shaping.  (b) Fourier transformation of time-domain oscillations
with chirality assigned peaks.}
\label{typical-nomask}
\end{figure}
%
%
\begin{figure} [tbp]
\includegraphics[scale=1.1]{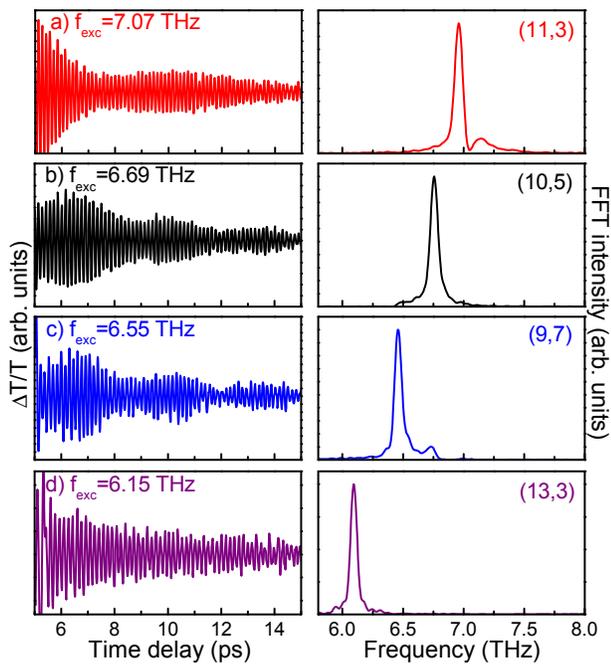}
\caption{(color online) (Left) Time-domain coherent RBM oscillations selectively
excited by multiple pulse trains via pulse-shaping with corresponding repetition
rates from 6.15-7.07~THz.  (Right) Fourier transformations of corresponding
oscillations, with their dominant nanotube chirality ($n$,$m$) indicated.}
\label{pulse-shaped-CP}
\end{figure}
%

Real-time observation of coherent RBM oscillations is possible without pulse-shaping
by employing standard femtosecond pump-probe spectroscopy.
\cite{Gambetta06.515,Lim06.2696,Lim07.306} Figure \ref{typical-nomask}(a) shows
transmission modulations of the probe beam induced by coherent lattice modulations,
which were generated by pump pulses with a
pulse width of 50 fs and a central wavelength of 800 nm (1.55~eV). The time-domain
beating profiles reflect the simultaneous generation of several RBM frequencies from
nanotubes in the ensemble with different chiralities, which are
clearly seen in Fig.~\ref{typical-nomask}(b) with the Fourier-transformation of
the time-domain data.  Although resonance conditions and mode frequencies lead to
the assignment of chiralities to their corresponding peaks,\cite{Lim06.2696} obtaining
detailed information on dynamical quantities such as the phase information of
phonon oscillations becomes rather challenging.  Additionally, if adjacent phonon
modes overlap in the spectral domain, this can lead to peak distortions.

However, by introducing pulse-shaping, multiple pulses with different repetition
rates are used to excite RBM oscillations, and as shown in
Figs.~\ref{pulse-shaped-CP}(a)-\ref{pulse-shaped-CP}(d), chirality selectivity
was successfully obtained. With the appropriate repetition rate of the pulse
trains, a single, specific chirality dominantly contributes to the signal, while
other nanotubes are suppressed.  For example, by choosing a pump repetition
rate of 7.07~THz, we can selectively excite only the $(11,3)$ nanotubes, as seen
in Fig.~\ref{pulse-shaped-CP}(a).  Similarly, with a pump repetition rate
of 6.69~THz, the $(10,5)$ nanotubes are selectively excited, as seen in
Fig.~\ref{pulse-shaped-CP}(b).  The accuracy of selectivity depends on the
number of pulses in the tailored pulse train as well as the
distribution of chiralities in the nanotubes ensemble. Furthermore, selective
excitation of a specific chirality also requires the pump energy
to be resonant with the corresponding $E_{22}$ transition for each
chirality-specific nanotubes.
%
\begin{figure} [tbp]
\includegraphics[scale=0.85]{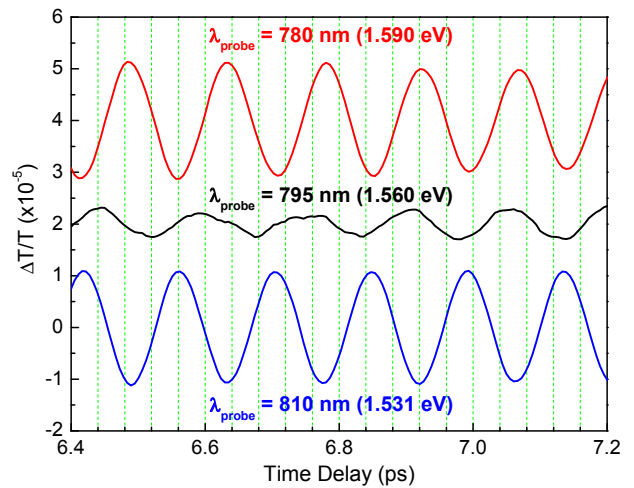}
\caption{(color online) Differential transmission as a function of time delay
at probe wavelengths of 780~nm, 795~nm, and 810~nm for the selective RBM excitation
of the (11,3) nanotubes.  There is a $\pi$ phase shift between the 780~nm and
810~nm data.  These three wavelengths (from the top to the bottom of the figure)
correspond to photon energies above, at and below the energy of the second
exciton resonance, respectively, of (11,3) nanotubes.}
\label{above-on-below}
\end{figure}

The ability to excite single-chirality nanotubes allows us to perform detailed
studies of excited states of single-walled carbon nanotubes.  For example, by
placing a series of 10-nm band pass filters in the probe path before the
detector, we can measure the wavelength-dependence of RBM-induced transmission
changes in order to understand exactly how the tube diameter changes during
coherent phonon RBM oscillations and how the diameter change modifies the nanotube
band structure.  As seen in Fig.~\ref{above-on-below}, the differential
transmission is shown for three cases, from top to bottom, corresponding to
probe photon energies above-resonance, on-resonance, and below-resonance,
respectively, for selectively-excited (11,3) carbon nanotubes.
Although the transmission is strongly modulated at the RBM
frequency (7.07~THz) for all three cases, the amplitude and phase of oscillations
vary noticeably for varying probe wavelengths.  Specifically, the amplitude of
oscillations becomes minimal at resonance, and, in addition, there is clearly
a $\pi$-phase shift between the above- and below-resonance traces. Because the band
gap energy and diameter are inversely related to each other, and because it is
the RBM frequency at which the diameter is oscillating, we can conclude from
this data that the energy of the $E_{22}$ resonance is oscillating at the RBM
frequency. Namely, when the band gap is decreasing, absorption above
(below) resonance is decreasing (increasing), resulting in positive (negative)
differential transmission.
%
\begin{figure} [tbp]
\includegraphics[scale=.4]{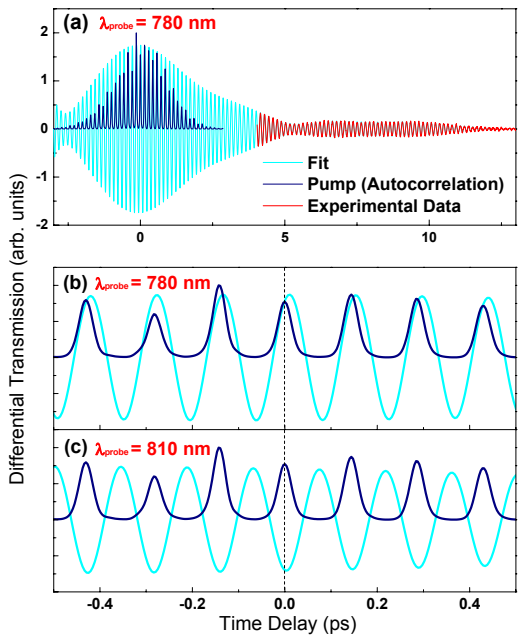}
\caption{(color online) (a) Differential transmission as a function of time delay
together with the pump pulse train (dark blue or dark gray) as well as fit
(light blue or light gray up to 4ps) to exponentially-decaying
sinusoidal oscillations. (b,c) Data near time zero for two wavelengths corresponding
to energies above and below the second exciton resonance, respectively, of
$(11,3)$ nanotubes.}
\label{initial-phase}
\end{figure}

We can also look at the short response to see how the diameter changes in
response to ultrafast excitation of electron-hole pairs by the pump pulse.
In Fig.~\ref{initial-phase}(a), we plot the differential transmission data
taken at 780~nm (1.59~eV) together with the pump pulse train, with time zero corresponding
to the center of the pulse train. Here we note that an increase in the absorption
corresponds to a decrease in the differential transmission.
Figures \ref{initial-phase}(b) and \ref{initial-phase}(c)
show data near time zero for two wavelengths corresponding to energies above and below
the second exciton resonance, respectively, of $(11,3)$ nanotubes.
The sign of the differential transmission oscillations in the first quarter-period,
where the time delay varies from $0.0\ \mbox{ps}$ to $0.07\ \mbox{ps}$,
is positive (negative) for the above (below) resonance probe, indicating that
there is an initial decrease (increase) in absorption for energies above (below)
resonance, demonstrating that the diameter of the nanotube initially expands,
taking into account the fact that the resonance energy is inversely related to
the diameter.  This initial expansion of the tube diameter is in agreement
with our theoretical predictions for the photoexcitation of coherent phonon
RBM oscillations by ultrafast laser pulses pumping near the $E_{22}$ transition
energy in mod~2 nanotubes [e.g., $(11,3)$ tubes].

\section{Theory}
\label{Theory section}

We have developed a microscopic theory for the generation of coherent phonons
in single-walled carbon nanotubes and their detection by means of coherent
phonon spectroscopy experiments. Our approach is based on obtaining equations
of motion for the coherent phonon amplitudes from the Heisenberg
equations of motion as described by Kuznetsov \textit{et al.} in
Ref. \onlinecite{kuznetsov94.3243}. In our theoretical model, we explicitly
incorporate the electronic energies and wavefunctions for the $\pi$ electrons,
the phonon dispersion relations and the corresponding phonon modes, the
electron-phonon interaction, the optical matrix elements, and the interaction
of carriers with a classical ultrafast laser pulse. For simplicity, and to
make the problem tractable, we neglect the many-body Coulomb interaction and
interactions with the surrounding liquid medium in the micelle-suspended
nanotube ensemble.

We are able to treat nanotubes of arbitrary chirality by exploiting all the
screw symmetry operations. This allows us to examine trends in the CP signal
strength within and between nanotube families. In addition, we gain something in
our conceptional understanding by deriving a simple driven oscillator equation
for the coherent phonon amplitudes where the driving function depends explicitly
on the time-dependent photoexcited carrier distribution functions. In the limit
where we ignore Coulomb interactions, the driven oscillator equation for the
coherent phonon amplitudes turns out to be exact.\cite{kuznetsov94.3243}

\subsection{Electron Hamiltonian}
\label{Electron hamiltonian subsection}

\begin{figure} [tbp]
\includegraphics[scale=.75]{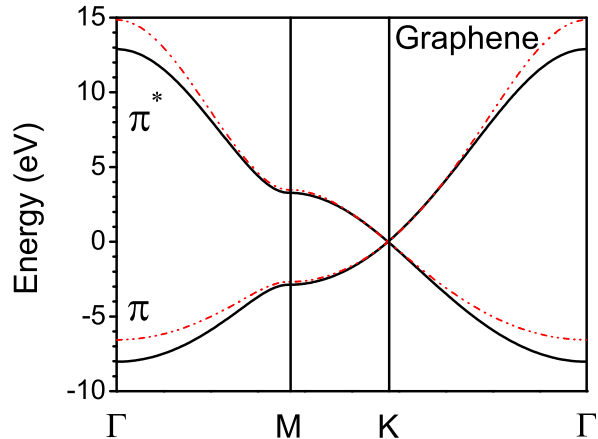}
\caption{
(color online) Extended tight-binding energy dispersion relations for the
bonding $\pi$ and anti-bonding $\pi^{*}$ bands in graphene along high symmetry
directions in the Brilloin zone are shown as solid black lines.
Energy dispersion relations in the nearest neighbor tight-binding model
described in Ref.~\onlinecite{Saito2003} are shown as dash-doted red lines.
}
\label{Graphene Pi bands figure}
\end{figure}

We treat carbon nanotube $\pi$ and $\pi^*$ electronic states in the extended
tight-binding (ETB) formalism of Porezag {\it et al}.\cite{Porezag95.12947}
In the ETB model, the tight-binding Hamiltonian and overlap matrix elements
between $\pi$ orbitals on different carbon atoms are functions of the interatomic
distance. The position-dependent Hamiltonian and overlap matrix elements are
obtained from a parametrization of density-functional (DFT) results in the
local-density approximation (LDA) using a local orbital basis set as described
in Ref.~\onlinecite{Porezag95.12947}. Our computed energy
dispersion relations for the bonding $\pi$ and anti-bonding $\pi^{*}$ bands in
graphene are plotted in Fig.~\ref{Graphene Pi bands figure} along high symmetry
directions in the hexagonal two-dimensional Brillouin zone. For comparison, we
also plot the graphene energy dispersion relations obtained from the simple
tight-binding model (STB), described in Ref.~\onlinecite{Saito2003}, in which
only nearest neighbor Hamiltonian and overlap matrix elements are considered.
In the STB model, the values of the nearest neighbor Hamiltonian and overlap
matrix elements are $-3.033 \ \mbox{eV}$ and $0.129$,
respectively.\cite{Saito2003} The ETB and STB models agree with each other
near the K and K' points in the Brillouin zone and in carbon nanotubes these
are the states that give rise to the low lying conduction and valence subbands
that we are interested in.
%
%

In a carbon nanotube with chiral indices $(n,m)$, a translational unit cell
can be found such that the atomic structure repeats itself after translation of
the tube by a translational vector $\textbf{T}$ parallel to the tube
axis.\cite{Saito2003} The resulting Brillouin zone is one-dimensional
with $\arrowvert k \arrowvert \leq \pi / T$ and the number of two-atom
hexagonal cells in each translational unit cell is \cite{Saito2003}
\begin{equation}
N_{\mbox{hex}} = \frac{2 (n^2 + n m + m^2)}{\gcd(2n+m,2m+n)}
\label{Nhex equation}
\end{equation}
where $\gcd(i,j)$ is the greatest common divisor of integers $i$ and $j$.
If we only make use of translational symmetry in formulating the electronic
problem, the resulting size of the Hamiltonian and overlap matrices is
$2N_{\mbox{hex}} \times 2N_{\mbox{hex}}$ if we retain one $\pi$ orbital per site.

In practice, the size of the electronic Hamiltonian and overlap matrices obtained
using the nanotube translational unit cell can become prohibitively large, especially
for chiral nanotubes with $n > m \neq 0$. Fortunately, we can reduce the size of
the electronic problem by further exploiting the symmetry of the nanotube. As
pointed out in Ref.~\onlinecite{Popov04.17} a two-atom hexagonal unit cell in
graphene, with the two carbon atoms labeled $A$ and $B$, can be mapped onto the nanotube
by applying two different screw operations. If we make use of the screw symmetry
operations, we can block diagonalize the $2N_{\mbox{hex}} \times 2N_{\mbox{hex}}$
Hamiltonian and overlap matrices into $2 \times 2$ subblocks which we label $\mu$.
In carbon nanotubes, the subblock index $\mu$ labels the cutting lines in
the zone folding picture. For states near the Fermi energy, the cutting line
numbers $\mu$ have a nice geometrical interpretation as pointed out in
Ref. \onlinecite{Saito05.153413}.

Derivations for the Hamiltonian and overlap matrices for the electronic
states in a carbon nanotube are given in Appendix~\ref{Appendix A}.
We let $s = v,c$ label the valence and conduction band states, and the electronic
energies $E_{s \mu}(k)$ for a given cutting line are obtained by solving the matrix
eigenvalue problem in Eq.~(\ref{Electronic Matrix Eigenvalue Problem}).
The second quantized electron Hamiltonian is simply
\begin{equation}
\hat{H}_e = \sum_{s \mu k}
E_{s \mu}(k) \ c_{s \mu k}^{\dag} \ c_{s \mu k}^{}
\label{Second quantized electron Hamiltonian}
\end{equation}
where $c_{s \mu k}^{\dag}$ creates an electron in the state with energy $E_{s \mu}(k)$.

\subsection{Phonon Hamiltonian}

Following Jiang \textit{et al.} \cite{Jiang06.235434} and Lobo \textit{et al.} \cite{Lobo97.159},
we treat lattice dynamics in a carbon nanotube using a valence force field model.
In our force field model, we include bond stretching, in-plane bond bending,
out-of-plane bond bending, and bond twisting potentials. In constructing the
valence force field potentials, we take care that they satisfy the force
constant sum rule which requires that the force field potential energy remain
invariant under rigid translations and rotations
(see Ref.~\onlinecite{Madelung.1978}, p. 131). As pointed out in
Ref.~\onlinecite{Mahon04.075405}, a number of calculations in the literature use force
field models that violate the force constant sum rule and, as a result, fail to
reproduce the long wavelength flexure modes predicted by elasticity theory.

\begin{figure} [tbp]
\includegraphics[scale=.75]{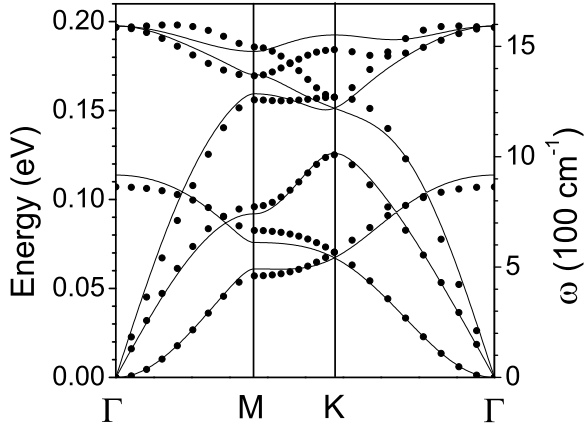}
\caption{
Graphene phonon dispersion relation. The solid circles are the valence force
field model of Jishi \textit{et al.} \cite{Jishi93.77} and the solid lines are the
best fit results for our seven parameter valence force field model.
}
\label{Graphene phonon dispersion figure}
\end{figure}

Our valence force field model as described in Appendix~\ref{Appendix B}
has seven force constants, four due to bond stretching interactions out to fourth nearest
neighbor shells and one each from the remaining three interactions. To determine these
seven force constants, we fit our model results for planar graphene to the model of
Jishi \textit{et al.} \cite{Jishi93.77} Our best fit dispersion relations are shown in
Fig.~\ref{Graphene phonon dispersion figure} as solid lines while the results of
Jishi \textit{et al.}~are shown as solid circles.

We should point out that since our force field model contains force constants that
are independent of the density of photoexcited carriers, it cannot describe phonon
softening which is observed at high values of the laser fluence. However, a rather high
value of the laser fluence is generally needed to generate a high density of
photoexcited carriers. In the case of metallic nanotubes, the chirality dependent frequency
shift of the RBM and G modes has been studied by Sasaki \textit{et al.}
in Refs.~\onlinecite{Sasaki08.235405} and \onlinecite{Sasaki08.245441}
as a function of the Fermi energy. In the case of the RBM mode they find that armchair
nanotubes do not exhibit any frequency shift while zigzag nanotubes exhibit phonon softening.
\cite{Sasaki08.235405}

The phonon energies and corresponding mode displacement vectors are obtained
by diagonalizing the dynamical matrix. In graphene, there are two atoms per
unit cell giving rise to the six phonon modes shown in Fig.~\ref{Graphene phonon dispersion figure}.
In a carbon nanotube, the size of the dynamical matrix is
$6 N_{\mbox{hex}} \times 6 N_{\mbox{hex}}$. By making use of the nanoutube screw symmetry
operations, we can block diagonalize the dynamical matrix into $6 \times 6$ subblocks
which we label $\nu = 0 \ldots N_{\mbox{hex}}-1$. Again, the subblock index $\nu$ labels
the cutting lines. In a carbon nanotube, the phonon energies $\hbar \omega_{\beta \nu}(q)$
are obtained from solving the dynamical matrix eigenvalue problem in
Eq.~(\ref{dynamical equations of motion}) where $\nu$ is the cutting line index and
$\beta = 1 \ldots 6$ labels the six modes associated with each cutting line. The
phonon wavevector $q$ is defined on a one-dimensional Brillouin zone given by
$\arrowvert q \arrowvert \le \pi / T$.

The second quantized phonon Hamiltonian is given by
\begin{equation}
\hat{H}_{ph} = \sum_{\beta \nu q}
\hbar \omega_{\beta \nu}(q) \ b_{\beta \nu q}^{\dag} \ b_{\beta \nu q}^{}
\label{Second quantized phonon Hamiltonian}
\end{equation}
where $b_{\beta \nu q}^{\dag}$ creates a phonon in a state with energy
$\hbar \omega_{\beta \nu}(q)$.

%

\subsection{Electron-phonon coupling}

The position-dependent single-electron Hamiltonian for a single-walled carbon
nanotube is given by
\begin{equation}
H(\textbf{r}) = T_0 + \sum_{r\textbf{J}}
v_c(\textbf{r} - (\textbf{R}_{r\textbf{J}}+\textbf{U}_{r\textbf{J}}))
\label{Single electron Hamiltonian}
\end{equation}
where the kinetic energy is $T_0$ and $v_c(\textbf{r})$ is the carbon atom potential.
In Eq.~(\ref{Single electron Hamiltonian}) $\textbf{R}_{r\textbf{J}}$
is the equilibrium position of the $r$-th atom in the $\textbf{J}$-th two-atom unit cell
and $\textbf{U}_{r\textbf{J}}$ is the corresponding atomic displacement from equilibrium
as defined in Appendix~\ref{Appendix B}.

Expanding this Hamiltonian in a Taylor series to first order in the atomic
displacements, we obtain a position-dependent electron-phonon interaction Hamiltonian
\begin{equation}
H_{ep}(\textbf{r}) = - \ \sum_{r\textbf{J}}
\nabla v_c(\textbf{r}-\textbf{R}_{r\textbf{J}}) \cdot \textbf{U}_{r\textbf{J}}
\label{Electron phonon Hamiltonian}
\end{equation}

Starting from the classical symmetry-adapted atomic displacement
in Eq.~(\ref{atomic displacement vectors}) of Appendix~\ref{Appendix B},
we make the transition to quantum mechanics and define a
symmetry-adapted second-quantized phonon displacement operator
\begin{eqnarray}
\nonumber
\hat{\textbf{U}}_{r\textbf{J}} &=& \sum_{\beta \nu q}
A_{\beta,\nu}(q) \
S(\theta_\textbf{J}) \
\hat{\textbf{e}}_r(\beta, q,\nu)
\\ &\times&
\ e^{ i \phi_{\textbf{J}}(q,\nu) } \
\left( b^{}_{\beta \nu q} + b^{\dag}_{\beta \nu, -q} \right)
\label{phonon displacement operator}
\end{eqnarray}
with the quantized phonon amplitudes
\begin{equation}
A_{\beta,\nu}(q) = \frac{1}{2} \ \frac{\hbar}{\sqrt{M N_{\mbox{hex}} \ \hbar \omega_{\beta \nu}(q)}}
\label{quantized phonon amplitude}
\end{equation}
Using the phonon displacement operator in Eq.~(\ref{phonon displacement operator})
we arrive at the desired second-quantized electron-phonon Hamiltonian
\begin{eqnarray}
\nonumber
\hat{H}_{ep} &=& \sum_{s s' \beta} \sum_{\mu \nu} \sum_{k q} \
M_{s, s', \beta}^{\mu,\nu} (k,q)
\\ &\times& \
c^{\dag}_{s',\mu+\nu,k+q} \ c^{}_{s \mu k} \
\left( b^{}_{\beta \nu q} + b^{\dag}_{\beta \nu, -q} \right)
\label{second quantized Electron phonon Hamiltonian}
\end{eqnarray}
The first term in the electron-phonon
Hamiltonian, Eq.~(\ref{second quantized Electron phonon Hamiltonian}),
is proportional to the operator
$c^{\dag}_{s',\mu+\nu,k+q} \ c^{}_{s \mu k} \ b^{}_{\beta \nu q}$.
This term describes a phonon absorption process in which an electron with
energy $E_{s \mu}(k)$ absorbs a phonon with energy
$\hbar \omega_{\beta \nu}(q)$ and then scatters into the state with
energy $E_{s',\mu+\nu}(k+q)$. Likewise, the second term in
Eq.~(\ref{second quantized Electron phonon Hamiltonian}) describes
a phonon emission process.

Details concerning the evaluation of the electron-phonon interaction matrix
elements $M_{s, s', \beta}^{\mu,\nu} (k,q)$ are given in Appendix~\ref{Appendix C}.
In evaluating these matrix elements, we make explicit use of $2p_z$ atomic
wavefunctions and screened atomic potentials obtained from an
\textit{ab initio} calculation in graphene.

\subsection{Coherent phonon generation}

The second quantized phonon displacement operators in
Eq.~(\ref{phonon displacement operator}) are defined in terms of a sum over
phonon modes of the second quantized operator $b^{}_{\beta \nu q} + b^{\dag}_{\beta \nu, -q}$.
For each phonon mode in the nanotube we are thus motivated to define a coherent
phonon amplitude given by\cite{kuznetsov94.3243}
\begin{equation}
\label{coherent phonon amplitude}
\nonumber
Q_{\beta \nu q}(t) = \langle b_{\beta \nu q}(t) + b^\dag_{\beta \nu,-q}(t) \rangle
\end{equation}
where $\langle \ \rangle$ denotes the statistical average. Equations of motion
for $Q_{\beta \nu q}(t)$ can be obtained from the phonon and electron-phonon Hamiltonians
in Eqs.~(\ref{Second quantized phonon Hamiltonian}) and
(\ref{second quantized Electron phonon Hamiltonian}). From the Heisenberg equation we obtain
\begin{eqnarray}
\label{Qmq(t) equation of motion}
\nonumber &&
\frac{\partial^2 Q_{\beta \nu q}(t)}{\partial t^2} + \omega^2_{\beta \nu}(q) \ Q_{\beta \nu q}(t) =
- \ \frac{2\omega_{\beta \nu}(q)}{\hbar}
\\ &&
\times \sum_{s s' \mu k}
M_{s, s', \beta}^{\mu,\nu} (k,-q)
\langle c^\dag_{s', \mu+\nu ,k-q}(t) \ c^{}_{s \mu k}(t) \rangle
\end{eqnarray}

In coherent phonon spectroscopy, we assume that the optical pulse and hence the
distribution of photoexcited carriers is spatially uniform over the nanotube. In this
case the electronic density matrix is diagonal and can be expressed as
\begin{equation}
\label{diagonal density matrix}
\langle c^\dag_{s', \mu+\nu ,k-q}(t) \ c^{}_{s \mu k}(t) \rangle =
\delta_{s,s'} \ \delta_{\nu,0} \ \delta_{q,0} \ f_{s \mu}(k,t)
\end{equation}
where $f_{s \mu}(k,t)$ is the electron distribution function in subband $s \mu$ with
wavevector $k$.\cite{kuznetsov94.3243}

The only coherent phonon modes that are excited are the $\nu = q = 0$ modes whose
amplitudes satisfy a driven oscillator equation
\begin{equation}
\label{Qm(t) equations of motion}
\frac{\partial^2 Q_{\beta}(t)}{\partial t^2} + \omega^2_\beta Q_{\beta}(t) = S_\beta(t)
\end{equation}
where $Q_\beta(t) \equiv Q_{\beta 0 0}(t)$ and $\omega_\beta \equiv \omega_{\beta,0}(q=0)$.
There is no damping term in Eq.~(\ref{Qm(t) equations of motion})
since anharmonic terms in the electron-phonon Hamiltonian are neglected.
We solve the driven oscillator equation subject to the initial
conditions $Q_\beta(0) = \dot{Q}_\beta(0) = 0$.  Taking the initial condition into
account, the driving function $S_\beta(t)$ is given by
\begin{equation}
\label{Sm(t) driving function}
S_\beta(t) = -\frac{2\omega_\beta}{\hbar} \sum_{s \mu k}
M^\beta_{s \mu}(k) \left( f_{s \mu}(k,t) - f^0_{s \mu}(k) \right)
\end{equation}
where $f_{s \mu}(k,t)$ are the time-dependent electron distribution
functions, $f^0_{s \mu}(k)$ are the initial equilibrium electron distribution
functions, and $M^\beta_{s \mu}(k) \equiv M^{\mu 0}_{s s \beta}(k,q=0)$.

The coherent phonon driving function $S_\beta(t)$ depends on the photoexcited
electron distribution functions. In principle, we could solve for the time-dependent
distribution functions in the Boltzmann equation formalism taking photogeneration
and relaxation effects into account. In CP spectroscopy, an ultrafast laser pulse
generates electron-hole pairs on a time scale short in comparison with the
coherent phonon period. In our experimental work, we typically use
50~fs ultrafast laser pulses to excite RBM coherent phonons with oscillation
periods of around 0.14~ps ($\hbar\omega_\beta \approx \mbox{30 meV}$ or $242 \ \mbox{cm}^{-1}$).

After photoexcitation the electron-hole pairs slowly scatter and recombine.
Jiang \textit{et al.} \cite{Jiang04.383} recently carried
out a study of the electron-phonon interaction and relaxation time in graphite and
found relaxation times on the order of a few picoseconds which is much slower
than either the ultrafast laser pulse or the RBM coherent phonon oscillation
period. The driving function $S_\beta(t)$ thus
rises sharply in a step-like fashion and then slowly vanishes as the distribution
functions $f_{s \mu}(k,t)$ return to $f^0_{s \mu}(k)$. The rapid initial jump in $S_\beta(t)$
gives rise to an oscillatory part of the coherent phonon amplitude $Q_\beta(t)$ at the
coherent phonon frequency $\omega_\beta$ while the slow subsequent decay of $S_\beta(t)$ gives
rise to a slowly varying background.  Since the observed CP signal is proportional
to the power spectrum of the oscillatory part of $Q_\beta(t)$, we choose to ignore
relaxation effects and retain only the rapidly varying photogeneration term in the
Boltzmann equation.  Neglecting carrier relaxation will have a negligible effect on
the computed CP signal since the relaxation time is much greater than the
coherent phonon period.

The photogeneration rate in the Boltzmann equation depends on the polarization
of the incident ultrafast laser pulse. Using an effective
mass model, Ajiki and Ando \cite{Ajiki94.349} showed that optical absorption in an
isolated single-walled carbon nanotube for polarization perpendicular
to the nanotube axis is almost perfectly suppressed by photo-induced charge
(the depolarization effect). Recently, Popov and Henrard \cite{Popov04.115407}
undertook a comparative study of the optical properties of
carbon nanotubes in orthogonal and nonorthogonal tight-binding models and
found that optical absorption due to light polarized parallel to the tube
axis ($z$ axis) is greater than absorption due to light polarized perpendicular
to the axis by about a factor of five. Consequently we confine our attention
to light polarized parallel to the tube axis only. We compute the photogeneration
rate in the electric dipole approximation using Fermi's golden rule. In the
case of parallel polarization, optical transitions can only occur between
states with the same angular momentum quantum number
$\mu$.\cite{Popov04.115407,Popov04.17}  For the photogeneration rate we find
\begin{eqnarray}
\nonumber &&
\left. \frac{\partial f_{s \mu}(k)}{\partial t} \right|_{\mbox{gen}} =
\frac{8 \pi^2 e^2 \ u(t)}{\hbar \ n_g^2 \ (\hbar\omega)^2}
\left(\frac{\hbar^2}{m_0} \right) \sum_{s'}
\arrowvert P^{\mu}_{s s'}(k) \arrowvert^2
\\ &&
\times \Big( f_{s'\mu}(k,t) - f_{s \mu}(k,t) \Big)
\ \delta \Big( \Delta E^{\mu}_{s s'}(k) - \hbar\omega \Big)
\label{photogeneration rate}
\end{eqnarray}
where $\Delta E^{\mu}_{s s'}(k) = \arrowvert E_{s \mu}(k) - E_{s' \mu}(k) \arrowvert$ are
the $k$ dependent transition energies, $\hbar\omega$ is the pump energy, $u(t)$ is
the time-dependent energy density of the pump pulse, $e$ is the electron charge,
$m_0$ is the free electron mass, and $n_g$ is the index of refraction in the
surrounding medium. The optical matrix element is given by
\begin{eqnarray}
\nonumber &&
P^{\mu}_{s s'}(k) = \frac{\hbar}{\sqrt{2 m_0}}
\sum_{r'} C^{*}_{r'}(s',\mu, k)
\\ &&
\times \sum_{r\textbf{J}}
\ C^{}_r(s, \mu, k) \ e^{i \phi_\textbf{J}(k,\mu)}
\ M_z(r',r\textbf{J})
\label{Pnn(k) equation}
\end{eqnarray}
where the sum over $r\textbf{J}$ is taken over fourth nearest neighbors of
the atom at $\textbf{R}_{r'\textbf{0}}$.
In Eq.~(\ref{Pnn(k) equation}), $C^{}_r(s, \mu, k)$ are the expansion
coefficients for the symmetry-adapted ETB wavefunctions obtained by
solving the matrix eigenvalue problem in Eq.~(\ref{Electronic Matrix Eigenvalue Problem})
of Appendix~\ref{Appendix A} and $\phi_\textbf{J}(k,\mu)$ is the phase
factor defined in Eq.~(\ref{phase factor equation}).

The $z$ components of the atomic dipole matrix elements
(which can be evaluated analytically) are given by
\begin{equation}
M_z = \int d\textbf{r} \
\varphi^{*}_{r'\textbf{0}}(\textbf{r}-\textbf{R}_{r'\textbf{0}}) \
\frac{\partial}{\partial z} \
\varphi_{r\textbf{J}}(\textbf{r}-\textbf{R}_{r\textbf{J}})
\label{Mz equation}
\end{equation}
where the $2p_z$ orbitals $\varphi_{r\textbf{J}}$ are defined in
Eq.~(\ref{pi orbital equation}). Note that the squared optical matrix
element $\arrowvert P^{\mu}_{s,s'}(k) \arrowvert^2$ has units of energy.
We point out that optical dipole matrix elements in the vicinity of
the $K$ point in both graphite and carbon nanotubes have been studied
previously in Ref.~\onlinecite{Gruneis03.165402}.

The pump energy density $u(t)$ is related to the fluence
$F = \int dt \ u(t) \ (c / n_g)$. To simplify our theoretical model, it is assumed that
the pump beam consists of a train of $N_{\mbox{pulse}}$ identical Gaussian pulses each
with an intensity full width at half maximum (FWHM) of $\tau_p$ and a Lorentzian spectral
lineshape with a FWHM of $\Gamma_p$. The Gaussian pulses are equally spaced in time with
the time interval between pulses being $T_{\mbox{pulse}}$. The peak intensity of the
first pulse is taken to occur at $t = 0$. To account for spectral broadening of the
laser pulses we replace the delta function in Eq.~(\ref{photogeneration rate}) with\cite{Chuang}
\begin{equation}
\delta(\Delta E-\hbar\omega) \rightarrow
\frac{\Gamma_p /(2\pi)} {{(\Delta E-\hbar\omega)^2+(\Gamma_p/2)^2}}
\label{delta function broadening}
\end{equation}

From the coherent phonon amplitudes, the time-dependent macroscopic displacements of
each carbon atom in the nanotube can be obtained by averaging
Eq.~(\ref{phonon displacement operator}). Thus
\begin{equation}
\label{coherent phonon displacement}
\textbf{U}_{r\textbf{J}}(t) = \sum_{\beta}
A_{\beta} \ S(\theta_\textbf{J}) \ \hat{\textbf{e}}_r^\beta \ Q_{\beta}(t)
\end{equation}
where $A_\beta \equiv A_{\beta,0}(0)$ and
$\hat{\textbf{e}}_r^\beta \equiv \hat{\textbf{e}}_r(\beta,0,0)$.

It is apparent that only four coherent phonon modes can be
excited in a carbon nanotube regardless of the chirality. Since $\nu = q = 0$ the
CP active mode frequencies and polarization vectors are found by diagonalizing a
single $6 \times 6$ dynamical matrix in Eq.~(\ref{dynamical equations of motion}).
Two of the six mode frequencies $\omega_\beta$ are zero and these phonons are not
excited since the driving term $S_\beta(t)$ vanishes as can be seen in Eq.~(\ref{Sm(t) driving function}).
Of the remaining four modes, the one with the lowest energy is the radial breathing mode.

Coherent acoustic phonon modes whose energies vanish at $q = 0$ cannot be excited in an
infinitely long carbon nanotube under conditions of uniform illumination by the pulse laser.
If, however, the electric field of the pump laser could be made to vary spatially along the
nanotube axis with a periodicity given by a real space wavevector $q_{\mbox{pulse}}$, it
would be possible to generate coherent acoustic phonons which would travel along the
nanotube at the acoustic sound speed. The generation of coherent acoustic phonons has
been demonstrated in semiconductor superlattices where the pump laser generates
carriers in the quantum wells thus giving rise to carrier distribution functions
having the periodicity of the superlattice.
\cite{Sanders01.235316,Sanders06.205303,Liu05.195335,Chern04.339}
In the case of coherent acoustic phonons in superlattices, the coherent phonon
lattice displacement satisfies a driven loaded string equation rather than a
driven harmonic oscillator equation.\cite{Sanders01.235316}

\subsection{Absorption spectrum}

In coherent phonon spectroscopy a probe pulse is used to measure the time-varying
absorption coefficient of the carbon nanotube. The time-dependent absorption
coefficient is given by\cite{Chuang, Bassani}
\begin{equation}
\label{absorption coefficient}
\alpha(\hbar\omega,t) =
\frac{\hbar\omega}{n_g \hbar c} \ \varepsilon_2(\hbar\omega,t)
\end{equation}
where $\varepsilon_2(\hbar\omega,t)$ is the imaginary part of the time-dependent
dielectric function evaluated at the probe photon energy $\hbar\omega$.

The imaginary part of the nanotube dielectric function is obtained from
Fermi's golden rule
\begin{eqnarray}
\label{imaginary dielectric function}
\nonumber &&
\varepsilon_2(\hbar\omega) =
\frac{8 \pi^2 e^2}{A_t (\hbar\omega)^2}
\left( \frac{\hbar^2}{m_0} \right) \sum_{s s' \mu }
\int \frac{dk}{\pi}
\ \arrowvert P^{\mu}_{s s'}(k) \arrowvert^2
\times
\\ &&
\Big( f_{s \mu}(k) - f_{s' \mu}(k) \Big)
\ \delta \Big( E_{s' \mu}(k) - E_{ s \mu }(k)- \hbar\omega \Big)
\end{eqnarray}
where $A_t=\pi (d_t/2)^2$ is the cross-sectional area of the tube and $d_t$ is
the nanotube diameter. In our model, we replace the delta function in
Eq.~(\ref{imaginary dielectric function}) with a broadened Lorentzian spectral
lineshape with a FWHM of $\Gamma_s$. The distribution function $f_{s \mu }(k)$
and bandstructure $E_{s \mu}(k)$ are time-dependent. The time-dependence of
$f_{s \mu}(k)$ comes from the Boltzmann carrier dynamics which can include
the photogeneration of electron-hole pairs as well as various carrier relaxation
effects. The time-dependence
of $E_{s \mu}(k)$ arises from variations in the carbon-carbon bond lengths due
to the coherent phonon induced atomic displacements $\textbf{U}_{r\textbf{J}}(t)$
given in Eq.~(\ref{coherent phonon displacement}). This time-dependent deformation
of the nanotube bond lengths alters the tight-binding Hamiltonian and overlap
matrix elements in the extended tight-binding model described in
Section~\ref{Electron hamiltonian subsection}. Note that to first order in the
lattice displacements the energies $E_{s \mu}(k)$ vary with time while the
tight-binding wavefunctions and optical matrix elements $P^{\mu}_{s s'}(k)$ do not.

\subsection{Coherent phonon spectrum}

In the coherent phonon spectroscopy experiments described by
Lim \textit{et al.}~in Refs.~\onlinecite{Lim06.2696} and \onlinecite{Lim07.306}
single color pump-probe experiments are performed on an ensemble of nanotubes.
The excitation of coherent phonons by the pump modulates the optical properties
of the nanotubes and gives rise to a transient differential transmission signal.
After subtraction of a slowly varying background component, the coherent phonon
spectrum is obtained by taking the power spectrum of the time-dependent
differential transmission. In our model we simulate single-color pump-probe
experiments and take the theoretical CP signal to be proportional to the power
spectrum of the transient differential transmission after background subtraction.
We compute the power spectrum using the Lomb periodogram algorithm described
in Ref.~\onlinecite{Press}. We find that using the Lomb periodogram
to evaluate the power spectrum in our theoretical results is more convenient
than using fast fourier transform methods since it works well for data sets
whose size is not an integer power of two or whose data points are not evenly
spaced.

It is worth emphasizing that in CP spectroscopy we measure the
power spectrum of the time-dependent coherent phonon-modulated differential
transmission. Thus, as a function of the pump energy, the CP spectrum at the
coherent phonon frequency tracks the absolute value of the first derivative
of the static absorption coefficient. This is nicely illustrated in Fig. 6 of
Ref.~\onlinecite{Lim06.2696} where it is shown that an excitonic peak in the
absorption spectrum will give rise to a symmetric double peaked structure in the
CP power spectrum.

\section{Theoretical Results}
\label{Results section}
%
\begin{figure} [tbp]
\includegraphics[scale=.95]{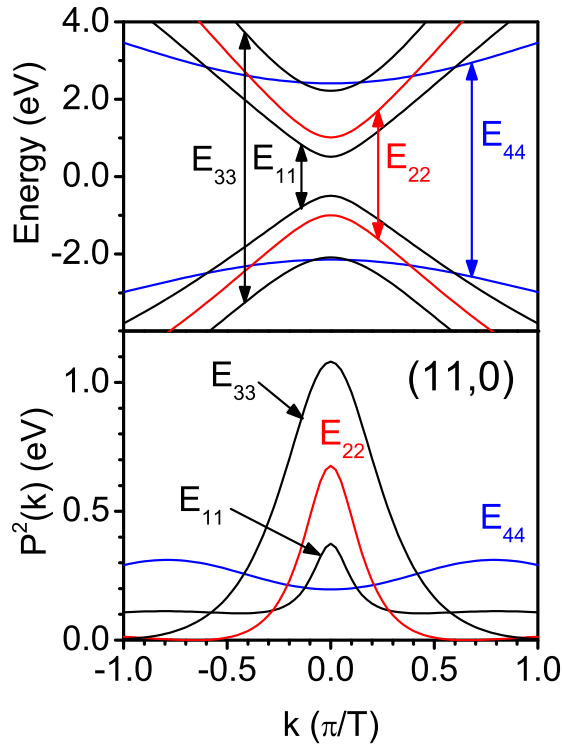}
\caption{
(color online) The upper panel show the four lowest lying electronic $\pi$
bands in the one-dimensional Brillouin zone for an $(11,0)$ zigzag nanotube in
the extended tight-binding (ETB) model. All the bands are doubly degenerate. The
allowed optical transition $E_{ii}$ for light polarized along the tube axis are
indicated. The lower panel shows the square of the optical matrix elements
defined in Eq.~\ref{Pnn(k) equation}
for each of the allowed optical transition.
}
\label{Enk 11 0 figure}
\end{figure}

To illustrate our theoretical model, we will discuss in some detail simulated
CP spectroscopy experiments in an undoped $(11,0)$ zigzag nanotube. This is a
mod~2 semiconducting nanotube with $\mod(n-m,3)=2$ belonging to the family of
nanotubes with $2 n + m = 22$. We choose this example since we have performed
CP spectroscopy for the $(11,0)$ nanotube (see Fig.~\ref{Family 22 and 25 CP comparison})
and because Lim \textit{et al.}~measured coherent lattice vibrations in a
micelle-suspended solution of carbon nanotubes with a diameter range
of 0.7 to 1.3~nm\cite{Lim06.2696,Lim07.306} and found strong CP signals due to
excitation of coherent RBM modes in this family of nanotubes.

\subsection{Bandstructure and absorption spectra}

The four lowest-lying one-dimensional electronic $\pi$ bands for the $(11,0)$
nanotube are shown in the upper panel of Fig.~\ref{Enk 11 0 figure}. In our model
we ignore structural optimization and assume that the carbon atoms lie on the
surface of the rolled up unreconstructed graphene cylinder. In an unreconstructed
zigzag nanotube the length of the translational unit cell is $T = \sqrt{3} \ a$
where $a = 2.49 \ {\AA}$ is the hexagonal lattice constant in graphene.\cite{Saito2003}
In Fig.~\ref{Enk 11 0 figure} the conduction bands have positive energy and the
valence bands have negative energy. Since the electronic problem has been reduced
to solving the $2 \times 2$ matrix eigenvalue problem in
Eq.~(\ref{Electronic Matrix Eigenvalue Problem}), the conduction bands in
Fig.~\ref{Enk 11 0 figure} with a given value of the angular quantum number $\mu$
can only mix with the valence band having the same value of $\mu$. The four bands
shown in the figure are doubly degenerate with two distinct values of $\mu$ giving
rise to the same band energies. The allowed optical transitions for $z$-polarized light
(with selection rule $\Delta \mu = 0$) are indicated by vertical arrows and are
labeled $E_{11} \ldots E_{44}$.

The lower panel of Fig.~\ref{Enk 11 0 figure} shows the square of the optical
matrix elements defined in Eq.~(\ref{Pnn(k) equation}) for each of the transitions
$E_{11} \ldots E_{44}$. For the $E_{11}$, $E_{22}$ and $E_{33}$ transitions
the squared optical matrix elements are strongly peaked at the van Hove singularity
at the direct band gaps. The size of this peak in the squared optical matrix elements
increases as the band gap moves away from the Fermi energy at $E = 0$. The absorption
for these three transitions are sharply peaked at the band edge due to the van Hove
singularity in the joint density of states as well as the peak in the squared
optical matrix elements that occurs there. For the $E_{44}$ transition, the conduction
and valence bands are very flat giving rise to an enhanced van Hove singularity
in the joint density of states while the squared optical matrix element is a
slowly varying function of $k$. For this transition, the peak in the absorption
spectrum is due almost entirely to the sharply peaked joint density of states.

As a check on our theoretically calculated squared optical matrix elements, we
compared our results with optical dipole matrix elements calculated independently
by Jiang \textit{et al.} in Ref.~\onlinecite{Jiang04.3169} for the metallic (5,5)
armchair and (6,0) zigzag tubes for light polarized parallel
to the tube axis. Our squared optical matrix elements are proportional
to the square of the dipole matrix elements shown in Figs.~2 and 3 in
Ref.~\onlinecite{Jiang04.3169}. We found excellent agreement between our
theoretical results and the corresponding results for the two tubes considered.

%
\begin{figure} [tbp]
\includegraphics[scale=.8]{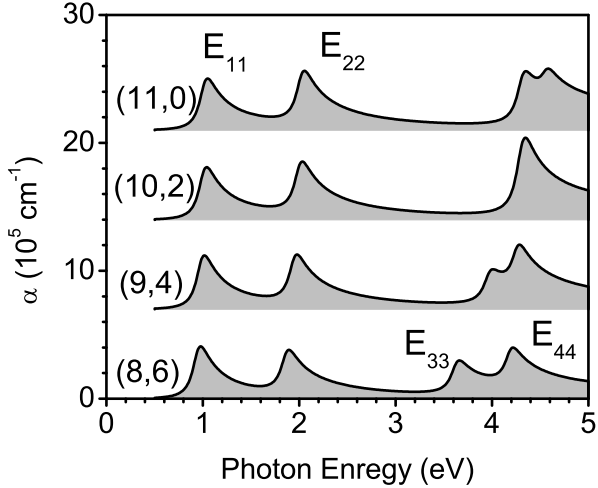}
\caption{
Computed absorption coefficient as a function of photon energy for all nanotubes
$(n,m)$ in the family for which $2n + m = 22$. The photons are assumed to be
linearly polarized with the electric polarization vector parallel to the tube
axis. Exciton effects are neglected.
}
\label{Family 22 alpha figure}
\end{figure}

With the electronic band structure and squared optical matrix elements shown in
Fig.~\ref{Enk 11 0 figure}, we can obtain the absorption coefficient of the $(11,0)$
nanotube using Eqs.~(\ref{absorption coefficient}) and (\ref{imaginary dielectric function})
and the carrier distribution functions.  The computed absorption coefficient for the
undoped $(11,0)$ nanotube in thermal equilibrium at room temperature is shown in
the upper curve of Fig.~\ref{Family 22 alpha figure} where the spectral FWHM
linewidth is taken to be $\Gamma_s = 0.15 \ \mbox{eV}$. Also shown in the figure are
the absorption spectra of the other members of Family 22, namely the
$(10,2)$, $(9,4)$ and $(8,6)$ nanotubes.

\subsection{Generation of coherent RBM phonons}
\begin{figure} [tbp]
\includegraphics[scale=.8]{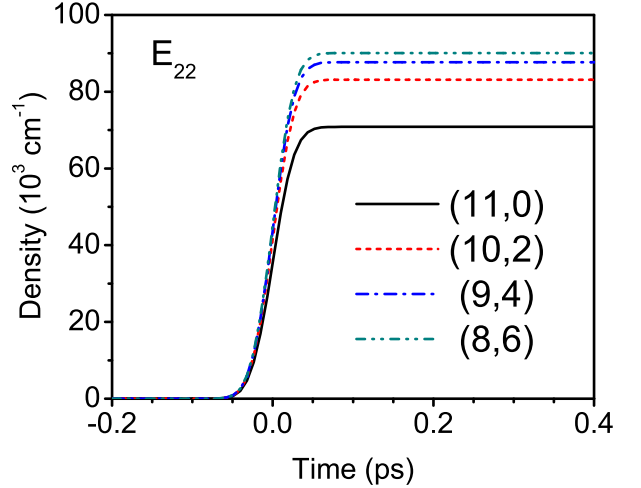}
\caption{
(color online)
Computed photoexcited carrier density per unit length for nanotubes in Family 22 after
photoexcitation by a 50 fs Gaussian laser pulse with a fluence of
$10^{-5} \ \mbox{J/cm}^2$ pumping at the peak of the $E_{22}$ features in
Fig.~\ref{Family 22 alpha figure}.
}
\label{Family 22 density figure}
\end{figure}
%
In a typical simulation, we excite coherent RBM phonons with a single
50~fs Gaussian laser pulse pumping at the peak of the broadened $E_{22}$ transitions shown in
Fig.~\ref{Family 22 alpha figure}. The pump fluence is taken to be
$10^{-5} \ \mbox{J/cm}^2$, the FWHM spectral linewidth is assumed to
be $\Gamma_p = 0.15 \ \mbox{eV}$, and the time scale is chosen so that the pump
reaches its peak intensity at $t = 0$. The pump energies $\hbar\omega$ at
the $E_{22}$ peaks are taken to be 2.05 eV for the $(11,0)$ nanotube, 2.04~eV for
$(10,2)$, 1.97~eV for $(9,4)$, and 1.89~eV for $(8,6)$. The photogenerated carrier
distribution functions are obtained from Eq.~(\ref{photogeneration rate}) and
the photoexcited carrier densities per unit length are shown in
Fig.~\ref{Family 22 density figure} for Family 22 nanotubes. The carrier
densities after photoexcitation all lie in the range from 70 to 90~cm$^{-1}$
increasing as we go from $(11,0)$ to $(8,6)$.
%
\begin{figure} [tbp]
\includegraphics[scale=.8]{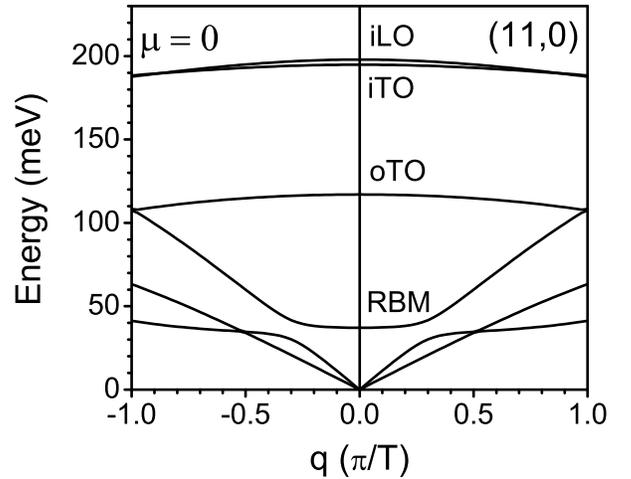}
\caption{
Phonon dispersion relations for $\mu = 0$ phonons in $(11,0)$
nanotubes. Only the four $q = 0$ modes with non-zero
frequency can be excited in coherent phonon spectroscopy.
}
\label{A1 phonon dispersion (11,0)}
\end{figure}

Using the photogenerated carrier densities $f_{s \mu}(k,t)$ we can obtain the
coherent phonon amplitudes $Q_\beta(t)$ by solving the equation of motion
(\ref{Qm(t) equations of motion}) with the driving function $S_\beta(t)$ given
in Eq.~(\ref{Sm(t) driving function}).  As we noted earlier, $\beta$ labels the
six coherent phonon modes for each value of $\mu$. Coherent phonon oscillations
can only be excited in the four $q = 0$ modes with non-zero frequency corresponding
to $\mu = 0$. The six $\mu = 0$ phonon dispersion curves are shown in
Fig.~\ref{A1 phonon dispersion (11,0)} for the $(11,0)$ nanotube. At $q = 0$, there
are two acoustic modes with zero frequency. The one with the lower sound speed is
the twisting mode (TW) in which the A and B atom sublattices move in phase in the
circumferential direction. The mode with the higher sound speed is the
longitudinal acoustic mode (LA) in which the A and B atom sublattices move in
phase along the tube axis.  The remaining four $q = 0$ modes are the CP active modes.
The lowest CP active mode is the radial breathing mode at 37.1~meV (300~cm$^{-1}$)
in which all the atoms move in phase radially in and out. The next highest CP
active mode is the out of plane transverse optical mode (oTO) at 117~meV (944~cm$^{-1}$)
in which the A atom sublattice moves radially outward as the B atom sublattice
moves radially inward. There are two closely spaced high frequency modes. The
lower of these at 194.8~meV (1571~cm$^{-1}$) is the in-plane transverse optical
mode (iTO) in which the A and B atom sublattices move in the circumferential
direction out of phase with each other. The highest mode is the longitudinal
optical mode (LO) at 197.9~meV (1596~cm$^{-1}$) in which the
A and B atom sublattices move out of phase with each other along the tube axis.

In our model we neglect slow carrier relaxation effects and retain only the
photogeneration term in the Boltzmann equation. The net photogenerated
conduction band electron distribution function $f_{c \mu}(k)-f^0_{c \mu} (k)$ is
then equal to the net photogenerated hole distribution function for each value of $k$.
In this case, we can obtain a simplified expression for the coherent phonon driving
function which only involves the photogenerated conduction band electron distributions.
We find
\begin{equation}
S_\beta(t) = \sum_{\mu k} S^\beta_\mu(k)
\left( f_{c \mu}(k)-f^0_{c \mu} (k) \right)
\label{Simple Sm(t) equation}
\end{equation}
The driving function kernel $S^\beta_\mu(k)$ is given by
\begin{equation}
S^\beta_\mu(k) = -\frac{2 \ \omega_\beta}{\hbar}
\left( M^\beta_{c \mu}(k)-M^\beta_{v \mu}(k) \right)
\label{Simple driving function kernel}
\end{equation}
where $M^\beta_{s \mu}(k)$ ($s = v,c$) are the same matrix elements appearing in
Eq.~(\ref{Sm(t) driving function}).
%
\begin{figure} [tbp]
\includegraphics[scale=.77]{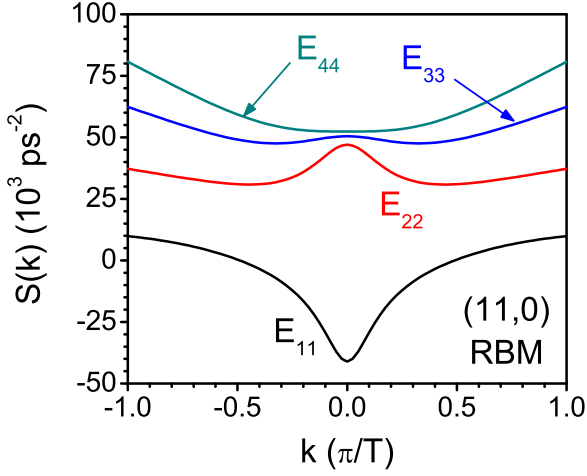}
\caption{
(color online)
Driving function kernel for RBM coherent phonons in the $(11,0)$ nanotube for
several optical transitions $E_{ii}$ in the impulsive excitation approximation.
}
\label{RBM driving function kernel}
\end{figure}
%
Each value of $\mu$ in the impulsive excitation model corresponds to a specific
optical transition $E_{ii}$. The $k$ dependence of $S_\mu(k)$ for the RBM phonon
in the $(11,0)$ nanotube is shown in Fig.~\ref{RBM driving function kernel} for the
first four optical transitions. For the RBM mode, we chose the unit mode
polarization vector to point radially outward so that positive values of $S(k)$
contribute to a radially outward directed driving term.

As can be seen in
Fig.~\ref{RBM driving function kernel} both positive and negative values of $S(k)$
are possible. If, for example, we were to pump near the $E_{11}$ band edge, the
electron distribution functions would be localized near $k = 0$ and we would get
negative values for $S(t)$. The signs of the driving function kernels near $k = 0$
for the $E_{ii}$ transitions are in agreement with other results reported in the
literature.\cite{Machon05.035416}
The sign of $S^\beta_\mu(k)$ is the negative of the sign of the electron-phonon matrix element
$M^\beta_{c \mu}(k)-M^\beta_{v \mu}(k)$ appearing in Eq.~(\ref{Simple driving function kernel}).
This matrix element has been obtained by Mach\'{o}n \textit{et al}. in
an \textit{ab initio} calculation reported in Ref.~\onlinecite{Machon05.035416}
for the lowest four optical transitions with light polarized parallel to the tube axis.
In Table I of Mach\'{o}n \textit{et al}.\cite{Machon05.035416} the sign of the band edge
electron-phonon matrix element for $E_{11}$ in the $(11,0)$ tube has a sign opposite to
that of the higher lying transitions in agreement with the results shown in
Fig~\ref{RBM driving function kernel}.

%
\begin{figure} [tbp]
\includegraphics[scale=.9]{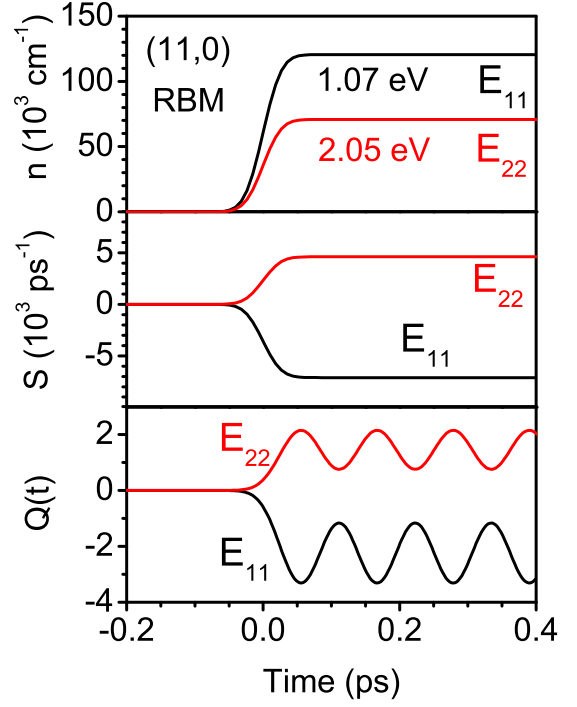}
\caption{
(color online)
Coherent phonon generation in $(11,0)$ nanotubes by photoexcitation at the
$E_{11}$ and $E_{22}$ transition energies. The upper panel shows the density of
photoexcited electron-hole pairs per unit length, the middle panel shows
the coherent phonon driving function, and the bottom panel shows the
RBM coherent phonon amplitude.
}
\label{E11 E22 CP comparison}
\end{figure}
%
In Fig.~\ref{E11 E22 CP comparison} we plot the
photoexcited carrier density $n(t)$, the coherent phonon driving function $S(t)$,
and the coherent phonon amplitude $Q(t)$ for RBM coherent phonons in $(11,0)$ tubes
for 50~fs $z$-polarized laser pulses with photoexcitation energies of 1.07~eV and
2.05~eV. These correspond to the $E_{11}$ and $E_{22}$ absorption peaks seen in
Fig.~\ref{Family 22 alpha figure} respectively. The absorption peaks are
comparable for $E_{11}$ and $E_{22}$. However, for constant fluence, the number
of photoexcited carriers $n \propto \alpha / \hbar\omega$
(see Ref.~\onlinecite{Chuang}, p. 341) and so the number of photoexcited
carriers is larger for the $E_{11}$ transition primarily as a result of the
smaller transition energy. As expected from Fig.~\ref{RBM driving function kernel},
the coherent phonon driving functions $S(t)$ and amplitudes $Q(t)$ have different signs
in the two cases. This means that for photoexcitation at the $E_{11}$ transition
in mod~2 nanotubes the tube diameter \textit{decreases} and oscillates about a
smaller equilibrium diameter while the opposite is true for photoexcitation at
the $E_{22}$ transition energy. As an aside, we note that in mod~1 tubes
the predicted $E_{22}$ and $E_{11}$ photoexcited diameter oscillations
have the opposite initial phase relative to the corresponding mod~2
oscillations as will be show in Section~\ref{CP detection in mod 1 tubes}.
%
\begin{figure} [tbp]
\includegraphics[scale=.9]{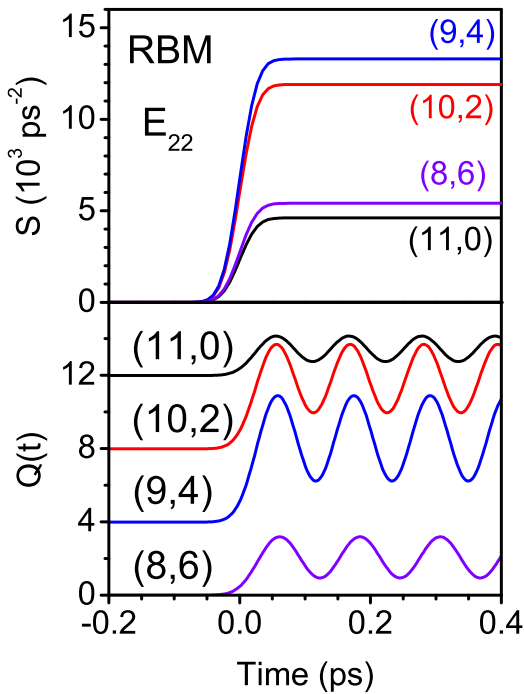}
\caption{
(color online)
Upper panel: Coherent phonon driving functions for RBM oscillations
in Family 22 nanotubes photoexcited by 50 fs pumps at the broadened $E_{22}$
absorption peaks in Fig.~\ref{Family 22 alpha figure}. Lower panel:
Corresponding coherent phonon amplitudes. Curves are offset for clarity.
}
\label{Family 22 RBM S(t) and Q(t)}
\end{figure}

If we pump at the broadened $E_{22}$ absorption peaks with 50~fs pulses of the
same spectral width for each nanotube in Family 22, we obtain the driving
functions shown in the upper panel of Fig.~\ref{Family 22 RBM S(t) and Q(t)}.
In all cases, the driving functions are positive which implies that all the
nanotubes in Family 22 initially move radially outward in response to
photoexcitation by an ultrafast pump at the $E_{22}$ transition energy. The
dimensionless coherent phonon amplitudes $Q(t)$ for the radial breathing
mode obtained by solving the driven oscillator equation
(\ref{Qm(t) equations of motion}) are shown in the lower panel of
Fig.~\ref{Family 22 RBM S(t) and Q(t)} where the curves are offset for clarity.
As expected, the coherent phonon amplitudes oscillate at the RBM frequencies
about a new positive equilibrium point. We note that for the radial breathing mode
the coherent phonon amplitude is proportional to the differential change in
the tube diameter.

In Fig.~\ref{Family 22 RBM S(t) and Q(t)}, we see that the
magnitude of the coherent phonon oscillations depends on chirality. The size of the
jump in the driving function $S(t)$ due to photoexcitation, and therefore the
magnitude of the oscillations in $Q(t)$, should be roughly
proportional to the product of the electron-photon and electron-phonon matrix elements.
The electron-phonon interaction matrix element for $E_{22}$ transitions in mod~2 tubes
is largest in the zigzag nanotube limit \cite{Jiang05.205420} while the corresponding
electron-photon interaction matrix element is largest in the armchair
limit.\cite{Jiang04.3169} Thus we expect $S(t)$ and $Q(t)$ to have maxima
somewhere between these two limits.

\subsection{Resonant excitation of coherent phonons}
\begin{figure} [tbp]
\includegraphics[scale=.9]{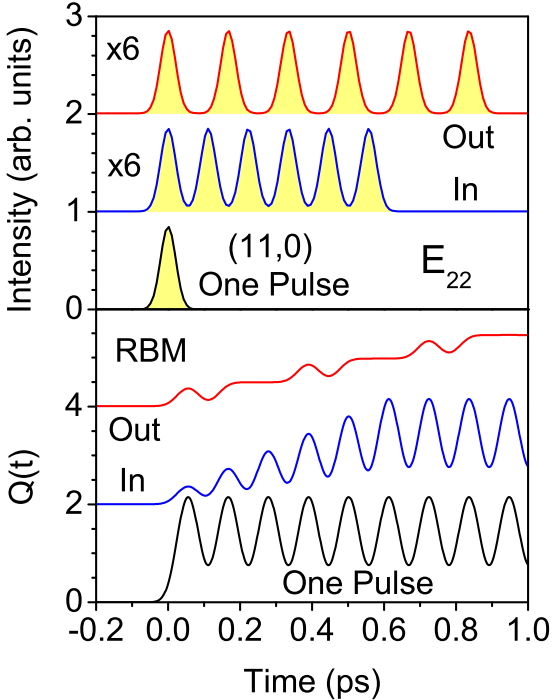}
\caption{
(color online) Upper panel: Pump laser intensity vs. time for a single 50 fs
Gaussian pulse and two trains of six 50 fs Gaussian pulses with repetition
periods in phase and 180 degrees out of phase with the $(11,0)$ RBM phonon
period. In all cases, we pump at the $(11,0)$ $E_{22}$ transition energy
(2.05~eV) with the same fluence. Note that the intensities of the two pulse trains are
magnified by a factor of six relative to the intensity of a single Guassian pulse.
Lower panel: Corresponding coherent phonon amplitudes for the $(11,0)$ RBM coherent
phonon oscillations.
}
\label{Pulse Train Figure (11,0)}
\end{figure}
%
In our experimental work, we resonantly excite coherent RBM phonons in
specific chirality nanotubes in a micelle suspended sample by using a
train of pump pulses with a repetition rate equal to the RBM period. To
illustrate resonant excitation of coherent phonons in nanotubes, we repeated our
simulations for the $(11,0)$ nanotube using a train of six Gaussian pulses pumping
at the $E_{22}$ transition energy (2.05~eV) where the pulse repetition period was in
phase and 180 degrees out of phase with the period of the RBM phonon.

The results are illustrated in Fig.~\ref{Pulse Train Figure (11,0)}. The
upper panel of Fig.~\ref{Pulse Train Figure (11,0)} shows the pump laser
intensity as a function of time for the in-phase and out-of-phase cases.
For comparison, we also plot the pump laser intensity for the single 50 fs Gaussian
pulse used in our earlier simulation. In all three cases the fluence is
taken to be $10^{-5}$J/cm$^2$
and hence the final density of photogenerated carriers were the same. We note that in the
figure, the intensities of the two pulse trains are multiplied by a factor of six
since the intensities of the Gaussian pulses scale inversely with the number of
pulses if the fluence is held constant.

The corresponding coherent phonon amplitudes for the RBM phonons are shown
in the lower panel of Fig.~\ref{Pulse Train Figure (11,0)}. For the single
Gaussian pulse, the coherent phonon amplitude is the same as that shown
in Fig.~\ref{Family 22 RBM S(t) and Q(t)} for the (11,0) nanotube. For the
in-phase case, the coherent phonon amplitude is magnified after each Gaussian
pulse and at long times is identical to the coherent phonon amplitude obtained
using the single Gaussian pulse. Since it is the long time behavior of the
oscillating part of the coherent phonon amplitude that determines the CP signal,
the two resulting CP spectra are also identical. When the Gaussian pulses
are 180 degrees out of phase, the oscillating part of the coherent phonon
amplitude and hence the resulting CP spectra are completely suppressed.

In Fig.~\ref{Pulse Train Figure (11,0)} the predicted
coherent phonon amplitudes at long times oscillate rapidly about the steady
state value $\bar{Q}_\beta = S_\beta(t \rightarrow \infty) / \omega_\beta^2$.
In real nanotubes the driving function $S_\beta(t)$ will slowly vanish as
the carriers recombine. In addition the coherent phonon amplitudes will slowly
decay on a time scale of tens of picoseconds as evidenced in
Fig.~\ref{pulse-shaped-CP}.

\subsection{Coherent phonon detection in mod 2 tubes}

The generation of coherent phonons results in periodic oscillations of the
carbon atoms which in turn modulate the optical properties of the nanotube.
These coherent phonon oscillations can be detected by measuring transient optical
properties in pump-probe experiments. We simulate single color pump-probe
measurements in which we pump with light linearly polarized along the tube axis
and measure the transient differential gain as a function of probe delay for a
probe pulse having the same energy and polarization as the pump. The coherent
phonon spectrum is obtained by scanning the pump-probe energy and in our
simulations, the pump fluence, duration, and FWHM spectral linewidth are assumed
to be constant as we vary the pump-probe energy.
%
\begin{figure} [tbp]
\includegraphics[scale=.75]{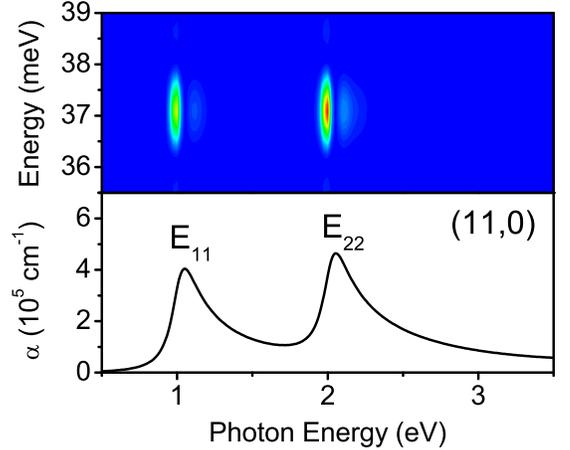}
\caption{
(color online)
The lower panel shows the absorption coefficient in (11,0) nanotubes as a
function of photon energy for light polarized along the tube axis. The upper
panel is a contour map showing the coherent phonon power spectrum as a
function of pump-probe energy and phonon frequency.
}
\label{CP spectrum for (11,0)}
\end{figure}

The coherent phonon (CP) spectrum for mod 2 $(11,0)$ nanotubes is shown in
Fig.~\ref{CP spectrum for (11,0)}. The bottom panel shows the absorption spectrum
for light linearly polarized along the tube axis assuming a FWHM linewidth of 0.15~eV.
The upper panel shows a contour map of the coherent phonon power spectrum as a
function of pump-probe energy and photon energy. The CP intensity is proportional
to the power spectrum and as we scan in photon energy two large peaks are observed
at the $E_{11}$ and $E_{22}$ transitions at a phonon energy near 37.1~meV (300~cm$^{-1}$)
which corresponds to the RBM frequency of the (11,0) nanotube. Comparing the upper and
lower panels of Fig.~\ref{CP spectrum for (11,0)}, we can verify our earlier assertion
that as we scan in pump energy the CP intensity at the RBM coherent phonon frequency
is proportional to the absolute value of the first
derivative of the absorption coefficient. We should point out that our theoretical
model does not include many-body Coulomb effects and so the position and shape of
the CP signal is due to modulation of the free carrier $E_{11}$ and $E_{22}$ transitions.
Thus in our free carrier model, we see an asymmetric double peak at each transition
with the stronger peak at low pump energy and the weaker peak at higher energy.
%
\begin{figure} [tbp]
\includegraphics[scale=1.0]{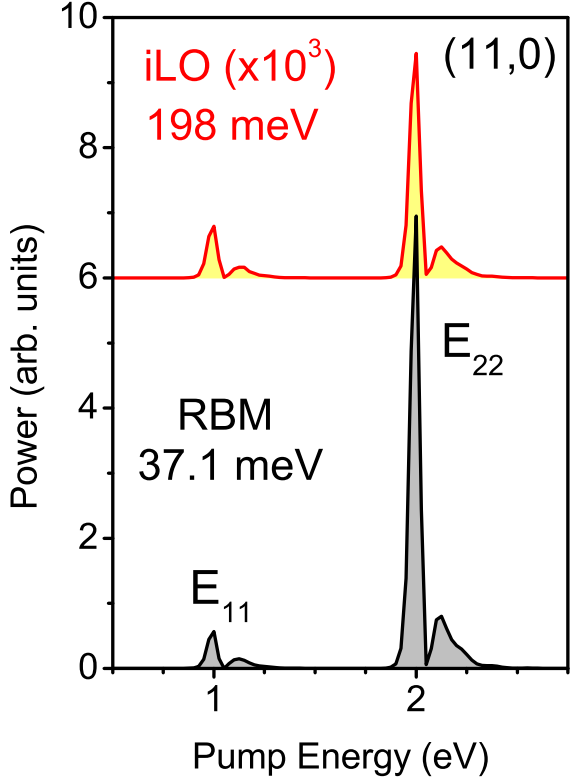}
\caption{
(color online)
Coherent phonon spectra for coherent RBM and LO phonons in $(11,0)$
nanotubes photoexcited by an ultrafast z-polarized 5 fs pump. The CP spectra
at the RBM and LO energies (37.1 and 198 meV) are plotted as a function of
pump-probe energy. Two features are seen near $E_{11}$ and $E_{22}$ transitions.
The LO curve is multiplied by 1000 and offset for clarity.
}
\label{Fast pulse CP spectum for (11,0)}
\end{figure}

In our simulations with 50~fs laser pulses, we
excite coherent RBM phonon modes since the pulse duration is much less than the
RBM phonon oscillation period. To excite the higher lying coherent phonon modes
it is necessary to use shorter laser pulses. Recently,
Gambetta \textit{et al.} \cite{Gambetta06.515} used sub-10-fs laser pulses to
excite the RBM and LO modes while excitation of the oTO and iTO modes were not
observed. To examine this case, we simulated CP spectroscopy in $(11,0)$ nanotubes
using short 5~fs laser pulses. In qualitative agreement with the measurements of
Gambetta \textit{et al.}, we find that the two strongest modes are the RBM and
LO modes while the strengths of the oTO and iTO modes are found to be negligible.
Our result for the iTO mode is consistent with the chirality dependent Raman
G-band intensity in which the iTO signal is absent in zigzag
nanotubes.\cite{Saito04.085312} The CP spectra for the RBM and LO modes are shown
in Fig.~\ref{Fast pulse CP spectum for (11,0)} where the differential gain power
spectra at the RBM and LO energies (37.1 and 198~meV, respectively) are plotted
as a function of pump-probe energy. Two strong features are seen near the
$E_{11}$ and $E_{22}$ transition energies. In this example, the two curves have
similar shapes but the LO CP spectrum (multiplied by a factor of 1000 in the figure)
is much weaker than the corresponding RBM spectrum. We note that shortening the
duration of the laser pulse from 50~fs to 5~fs enhances the strength of the
$E_{22}$ peak relative to the $E_{11}$ peak. This can best be seen by comparing
the bottom curves in Fig.~\ref{Fast pulse CP spectum for (11,0)} with the
bottom curve in Fig.~\ref{Mod 1 and Mod 2 CP comparison}.
%
\begin{figure} [tbp]
\includegraphics[scale=.75]{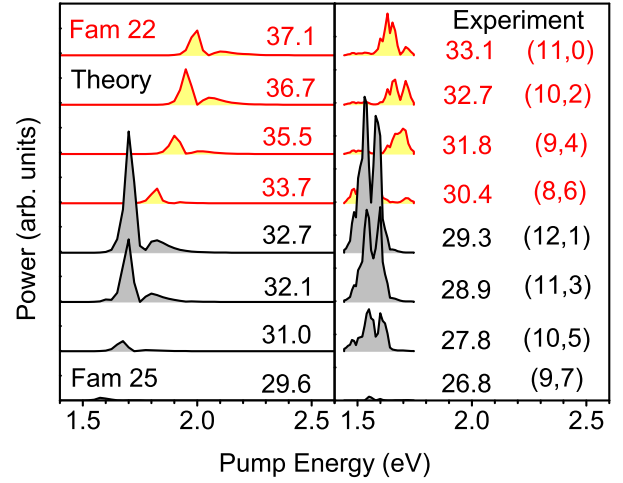}
\caption{
(color online) Coherent phonon intensity at the RBM frequency as a function of
pump-probe energy for several mod 2 semiconducting nanotubes at the $E_{22}$
transition. The experimental CP spectra are in the right panel and the simulated
CP spectra are in the left panel. The upper four curves in each panel are for
nanotubes in Family 22 and the lower four curves are for tubes in Family 25.
Each curve is labeled with the chirality $(n,m)$ and the RBM phonon energy in meV
and offset for clarity.
}
\label{Family 22 and 25 CP comparison}
\end{figure}

It is useful to examine trends in the CP spectra within and between mod~2
semiconducting nanotube families by plotting the theoretical CP intensity at
the RBM phonon frequency as a function of pump-probe energy. This is done in the
left panel of Fig.~\ref{Family 22 and 25 CP comparison} where we plot our
theoretical CP intensity at the RBM frequency as a function of pump-probe energy
for all nanotubes in Families 22 and 25. The curves for each nanotube are labeled with
the nanotube chirality $(n,m)$ and the RBM phonon energy in meV. In each nanotube,
we see peaks in the CP spectra corresponding to $E_{22}$ transitions. Within a
given family, the CP intensity tends to decrease as the chiral angle increases,
i.e., as the chirality goes from $(n,0)$ zigzag tubes to $(n,n)$ armchair tubes.
From Fig.~\ref{Family 22 and 25 CP comparison} we can also see that the theoretical
CP intensity increases as we go from Family 22 to Family 25.

The right panel of Fig.~\ref{Family 22 and 25 CP comparison} shows the corresponding
experimental CP spectra for the nanotubes in Families 22 and 25.  Comparing
experimental and theoretical curves in Fig.~\ref{Family 22 and 25 CP comparison}, we
see that our theory correctly predicts the overall trends in the CP intensities.
Since we are using pump probe methods to study an ensemble of micelle-suspended
nanotubes, the relative agreement between the theoretically calculated and
experimentally measured CP intensities suggests that nanotubes of
different chiralities in Families 22 and 25 in the micelle-suspended sample studied
are equally probable and that the measured CP signal strengths are an intrinsic
property of the tubes.

There are discrepancies in the predicted pump-probe energies of the peaks on the
order of $0.4 \ \mbox{eV}$ or less. A discrepancy of this size is expected since we
have not included many-body Coulomb interactions in our theoretical model. It is well
established that both the excitonic red shift and the self-energy blue shift are very
large in nanotubes, with the latter exceeding the former.
\cite{Ando05.777, Dresselhaus07.719, Ando97.1066, Jiang07.035407, Capaz06.121401, Dukovic05.2314}
We also note that the dielectric function of the surrounding medium also influences
the excitonic transition energies.\cite{Miyauchi07.394}
There are also differences between the theoretical and experimental CP lineshapes
in Fig.~\ref{Family 22 and 25 CP comparison}. If we compare theoretical and
experimental CP spectra for the (12,1) nanotube, we see that both exhibit a
double peaked structure. However, the lower energy theoretical peak is much stronger
than the higher energy peak whereas the two experimental peaks have comparable strength.
This discrepancy can be attributed to strong excitonic modification of the shape
of the nanotube absorption spectrum whose time-dependent modulation of the probe
pulse gives rise to the shape of the CP signal. The free carrier absorption edge
is highly asymmetric while the excitonic absorption spectrum exhibits a symmetric
peak at the exciton transition energy, thus accounting for the discrepancy.
Our theory qualitativley agrees with experiment, but to obtain quantitative
agreement, one must include details of the Coulomb interaction.

Our experimental and theoretical results are also
in qualitative agreement with the results of CP spectroscopy measurements
previously reported by Lim \textit{et al.}~in Ref.~\onlinecite{Lim06.2696}.
As these authors note, the tendency of the CP intensity to increase with
family index is in contrast to the situation in resonant Raman scattering where
the strength of the resonant Raman signal is observed to decrease as the family
index increases.

\subsection{Coherent phonon detection in mod 1 tubes}
\label{CP detection in mod 1 tubes}

\begin{figure} [tbp]
\includegraphics[scale=1.0]{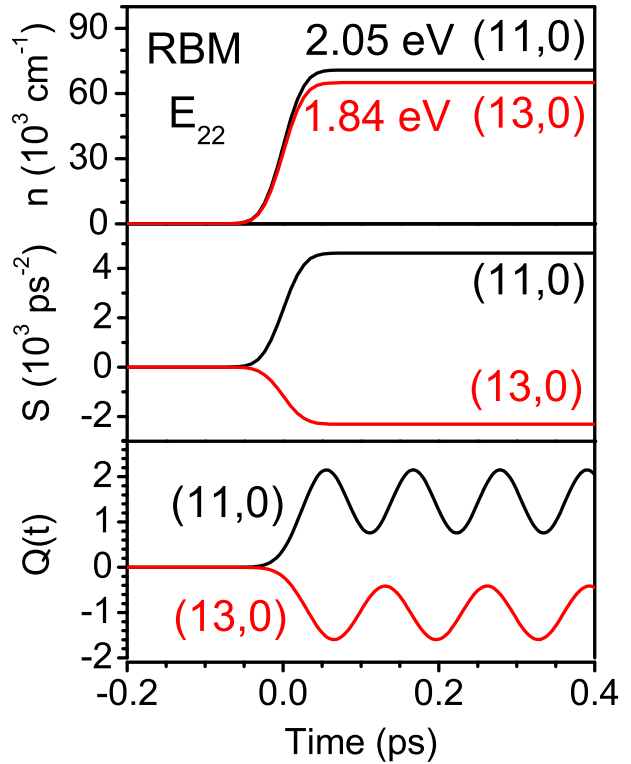}
\caption{
(color online)
Coherent phonon generation in $(11,0)$ mod 2 and $(13,0)$ mod 1 semiconducting
zigzag nanotubes by photoexcitation at the
$E_{22}$ transition energy. The upper panel shows the density of
photoexcited electron-hole pairs per unit length, the middle panel shows
the coherent phonon driving function, and the bottom panel shows the
RBM coherent phonon amplitude.
}
\label{Mod 1 and Mod 2 Three Panel Figure}
\end{figure}
%
It is useful to perform a comparison between mod~1 and mod~2 semiconducting
nanotubes with the same chiral angle. To this end, we compare the Family 22
$(11,0)$ mod~2 nanotubes with Family 26 $(13,0)$ mod~1 tubes. In both cases, the
pump lasers have the same fluence ($10^{-5} \ \mbox{J/cm}^2$), pulse duration
(50~fs), and spectral linewidth (0.15~eV), while the pump energies correspond
to the maxima in the broadened $E_{22}$ absorption features. For the $(11,0)$
tube we pump at 2.05~eV and for the $(13,0)$ tube we pump at 1.84~eV. The
time-dependent photogenerated carrier densities per unit length are shown in
the upper panel of Fig.~\ref{Mod 1 and Mod 2 Three Panel Figure}.  The middle
and lower panels show the coherent phonon driving functions and corresponding
coherent phonon amplitudes for RBM coherent phonons in the two cases. For $(11,0)$
mod~2 tubes, the coherent phonon driving function is {\em positive} while for
$(13,0)$ mod~1 tubes, the driving function is found to be {\em negative}.
%
\begin{figure} [tbp]
\includegraphics[scale=.77]{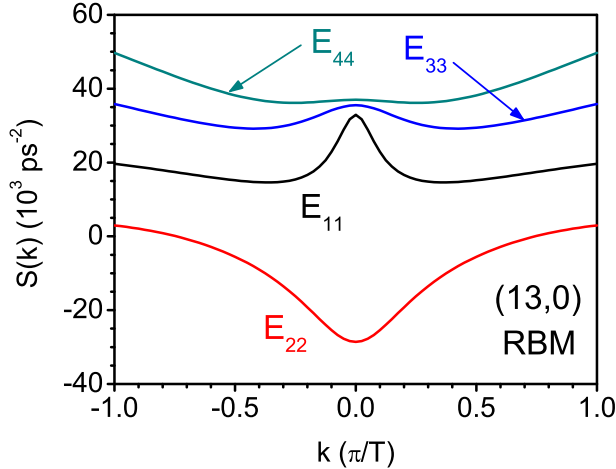}
\caption{
(color online)
Driving function kernel for RBM coherent phonons in the (13,0) nanotube for
several optical transitions $E_{ii}$ in the impulsive excitation approximation.
}
\label{RBM (13,0) driving function kernel}
\end{figure}
%
Some insight into this behavior can be obtained by examining the $k$-dependent
driving function kernel for coherent RBM phonons in $(13,0)$ nanotubes shown in
Fig.~\ref{RBM (13,0) driving function kernel}. Comparing
Fig.~\ref{RBM (13,0) driving function kernel}
with Fig.~\ref{RBM driving function kernel}, we see that the driving function
kernels for $E_{11}$ and $E_{22}$ transitions near $k = 0$ have opposite signs
which accounts for the sign change.

These results are supported by other studies reported in the literature.\cite{Jiang05.205420}
The difference in the sign of the $E_{22}$ driving function kernels at $k = 0$
for the zigzag mod 2 (11,0) tube shown in Fig.~\ref{RBM driving function kernel} and the
zigzag mod 1 (13,0) tube shown in Fig.~\ref{RBM (13,0) driving function kernel} is
due to a change in sign of the electron-phonon matrix element. This sign change
in the electron-phonon matrix element for tubes with different mod numbers was
also obtained independently by Jiang \textit{et al}. in Ref.~\onlinecite{Jiang05.205420}.
Figure 1(a) in Jiang \textit{et al}.\cite{Jiang05.205420} shows the electron-phonon
matrix element for coherent RBM phonons excited at the $E_{22}$ transition as a
function of the chiral angle $\theta$ for mod 1 and mod 2 tubes (the SII and SI curves in
Jiang's Fig.~1(a)). For zigzag tubes, $\theta = 0$ and from Jiang's Fig.~1(a) we
can see that the sign of the electron-phonon matrix element for the mod 2 (SI)
tube is positive while the sign of the electron-phonon matrix element for the mod 1 (SII)
tube is negative. This is consistent with the theoretical results shown in our
Figs.~\ref{RBM driving function kernel} and \ref{RBM (13,0) driving function kernel}.

%
\begin{figure} [tbp]
\includegraphics[scale=1.0]{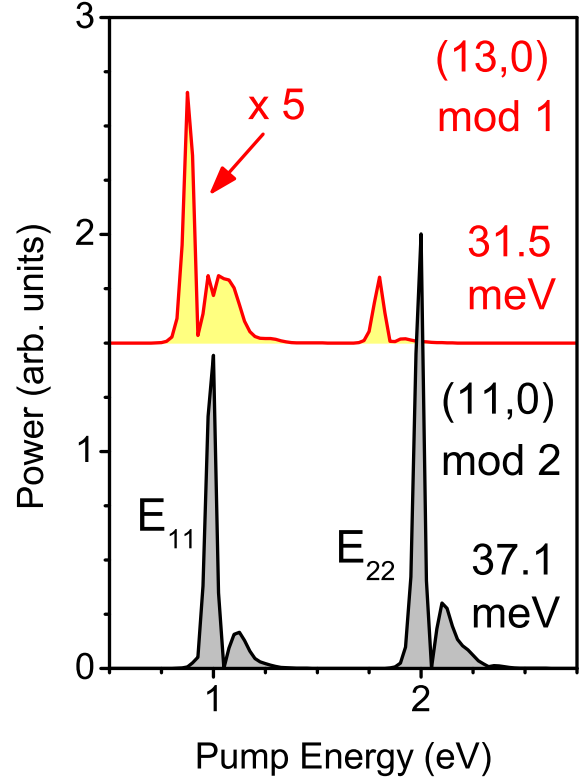}
\caption{
(color online)
Coherent phonon power spectra at the RBM energies in the $(13,0)$ mod 1 and
$(11,0)$ mod 2 semiconducting nanotubes as a function of pump-probe energy.
The mod 1 CP power spectrum is multiplied by a factor of 5 and offset for clarity.
}
\label{Mod 1 and Mod 2 CP comparison}
\end{figure}

We find that the CP intensity is very sensitive to the nanotube mod number.
In general, the CP intensity in mod~2 semiconducting nanotubes is much larger
than the CP intensity in mod~1 semiconducting tubes. This is illustrated
in Fig.~\ref{Mod 1 and Mod 2 CP comparison} where we plot the CP power spectra
as a function of pump-probe energy at the RBM frequencies for the zigzag $(13,0)$
mod~1 and (11,0) mod~2 semiconducting nanotubes. For the $(13,0)$ and $(11,0)$ tubes
the RBM phonon energies are 31.5 (254~cm$^{-1}$) and 37.1~meV (300~cm$^{-1}$),
respectively. Note that in
Fig.~\ref{Mod 1 and Mod 2 CP comparison} the mod~1 curve is multiplied by a
factor of 5. In general, we find that CP intensities in mod~2 tubes are
considerably larger than CP intensities in mod~1 tubes. This is consistent with
the experimental results of Lim \textit{et al.}~as reported in
Refs.~\onlinecite{Lim06.2696} and \onlinecite{Lim07.306}. We also find that in
mod~1 tubes, the $E_{11}$ feature is more pronounced than the $E_{22}$ feature
in contrast to what is seen in the mod~2 case.

\begin{figure} [tbp]
\includegraphics[scale=1.0]{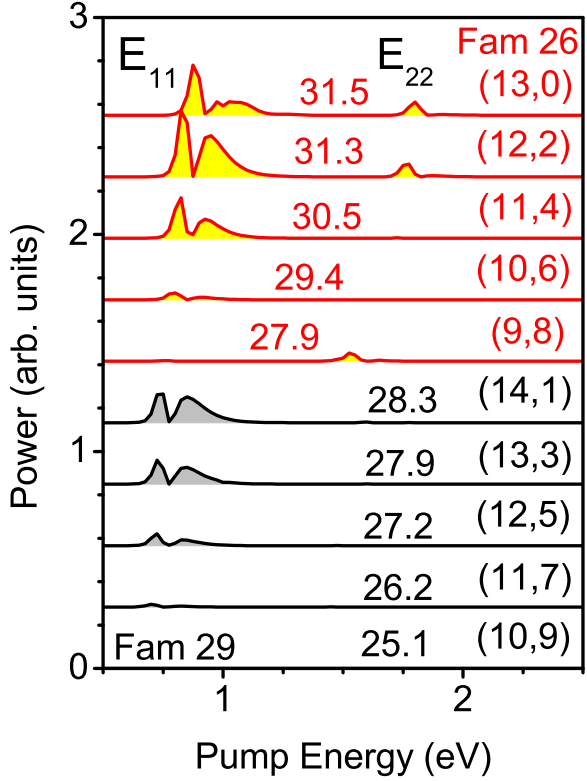}
\caption{
(color online)
Coherent phonon intensity at the RBM frequency as a function of pump-probe
energy for several mod 1 semiconducting nanotubes. The
upper five curves are for nanotubes in Family 26 and the lower five curves are
for tubes in Family 29. Each curve is labeled with the chirality $(n,m)$ and the
RBM phonon energy in meV. The curves are offset for clarity.
}
\label{Family 26 and 29 CP comparison}
\end{figure}
%
The CP intensities as a function of chirality for nanotubes in two mod~1 families
(Families 26 and 29) are shown in Fig.~\ref{Family 26 and 29 CP comparison}.
We find that in all cases the $E_{11}$ features in the mod~1 tubes are much stronger
than the $E_{22}$ features. Within a mod~1 family, the CP intensity of the $E_{11}$
feature is found to decrease with increasing chiral angle.  We also find that the
CP intensities decrease as the mod~1 family index increases.

\section{Summary}
\label{Summary section}

Using femtosecond pump-probe spectroscopy with pulse shaping techniques, we have
generated and detected coherent phonons in chirality-specific semiconducting
single-walled carbon nanotubes.  The signals were resonantly enhanced when the
pump photon energy coincides with an interband exciton resonance, and analysis
of such data provided a wealth of information on the chirality-dependence
of light absorption, phonon generation, and phonon-induced band structure modulations.

To explain our experimental results qualitatively and quantitatively, we have
developed a microscopic theory for the generation and detection of coherent phonons
in semiconducting single-walled carbon nanotubes via coherent
phonon spectroscopy. For extremely short laser
pulses, we find that the two strongest coherent phonon modes are the RBM and
LO modes. The CP spectrum of the LO mode is similar in shape to that of the
RBM mode but is found to be much weaker.

For the RBM modes, the CP intensity within a mod~2 family tends to decrease
with chiral angle, and the decrease in CP intensity with chiral angle is found
to be much more pronounced for the $E_{11}$ feature.  We also find that CP
intensities are considerably weaker in mod~1 families in comparison with
mod~2 families. In general, the $E_{22}$ CP intensities in mod~2 families are
stronger than the $E_{11}$ features.  For RBM modes in mod~1 tubes, the $E_{11}$
intensities are stronger than the $E_{22}$ intensities and tend to decrease with
increasing chiral angle within a given family. As the family index increases,
the $E_{11}$ CP intensity in mod~1 tubes decreases.

For mod~2 nanotubes, we predict that the tube diameter will initially increase
for $E_{22}$ photoexcitation and decrease for $E_{11}$ photoexcitation.
In mod~1 nanotubes, the opposite is precdicted to be the case, \textit{i.e.}
the tube diameter will initially decrease for $E_{22}$ photoexcitation
and increase for $E_{11}$ photoexcitation.

We compare our theoretical results with experimental CP spectra in mod~2 nanotubes
and find that our theoretical model correctly predicts the overall trends in
the relative strengths of the CP signal both within and between mod~2 families.
We find discrepancies between our theoretical predictions with regard to the
peak positions and lineshapes, which we attribute to Coulomb interaction effects
that are not included in our calculations.

For $(11,3)$ mod~2 nanotubes, we experimentally verified our theoretical
prediction that the diameter of $E_{22}$ photoexcited nanotubes initially
increases. However, we were unable to get a good sample for verifying our
related prediction that the diameter of $E_{22}$ photoexcited
mod~1 nanotubes initially decreases. This will be one of the goals of our
future studies.

\appendix
\section{Electronic States}
\label{Appendix A}

The tight-binding wavefunctions for the electronic states in carbon nanotubes
are expanded in terms of the symmetry-adapted basis functions
\begin{equation}
\arrowvert k,\mu \rangle = \sum_r C_r(k,\mu) \ \arrowvert k,\mu,r \rangle
\label{TB wavefunction}
\end{equation}
where $r=A,B$ labels the atoms in the two-atom unit cell, $C_r(k,\mu)$ are expansion
coefficients, and $\arrowvert k,\mu,r \rangle$ are symmetry-adapted basis functions.

The symmetry-adapted basis functions are linear combinations of localized atomic $\pi$ orbitals
\begin{equation}
\arrowvert k,\mu,r \rangle = \frac{1}{\sqrt{N}}
\sum_\textbf{J} e^{ i \hat{\textbf{k}}(k,\mu) \cdot \textbf{R}_{\textbf{J}} }
\ \arrowvert \textbf{J},r \rangle
\label{symmetry adapted basis functions}
\end{equation}
where $N$ is the number of two-atom unit cells in the system and $\arrowvert \textbf{J},r \rangle$
is a localized atomic $\pi$ orbital on atom $r$ in the two-atom unit cell at $\textbf{R}_{\textbf{J}}$.
The positions of the two-atom unit cells (in unrolled graphene $xy$ coordinates) are
\begin{equation}
\textbf{R}_{\textbf{J}} = j_1 \ \textbf{a}_1 +j_2 \ \textbf{a}_2
\label{graphene coordinates}
\end{equation}
where $\textbf{J} = (j_1,j_2)$ and $\textbf{a}_1 = (\frac{\sqrt{3}a}{2},\frac{a}{2})$
and $\textbf{a}_2 = (\frac{\sqrt{3}a}{2},-\frac{a}{2})$ are the graphene basis vectors.

The two dimensional wavevector $\hat{\textbf{k}}(k,\mu)$ appearing in the
symmetry-adapted basis function expansion in Eq.~(\ref{symmetry adapted basis functions})
is determined by imposing translational and rotational boundary conditions on the nanotube.
Imposing the translational boundary condition we have
\begin{equation}
\hat{\textbf{k}}(k,\mu) \cdot \textbf{T} = k \ T
\label{translational boundary condition}
\end{equation}
where $\textbf{T} = t_1 \ \textbf{a}_1 + t_2 \ \textbf{a}_2$ is the nanotube
translational vector which is parallel to the tube axis and has the length of the
translational unit cell. Explicit expressions for $t_1$ and $t_2$ in terms of the
chiral indices $n$ and $m$ can be found in Ref.~\onlinecite{Saito2003} and are
given by $t_1 = \frac{2 m + n}{d_R}$ and $t_2 = - \frac{2 n + m}{d_R}$, where
$d_R = \gcd(2 n + m, 2 m + n)$.  The rotational boundary condition is
\begin{equation}
\hat{\textbf{k}}(k,\mu) \cdot \textbf{C}_h = 2 \pi \mu
\label{rotational boundary condition}
\end{equation}
where $\textbf{C}_h = n \ \textbf{a}_1 +m \ \textbf{a}_2$ is the chiral vector in
unrolled graphene coordinates and $\mu = 0 \cdots N_{\mbox{hex}}-1$ is an angular momentum
quantum number that labels the cutting lines in the simple zone folding picture.\cite{Saito2003}

From Eqs.~(\ref{TB wavefunction}-\ref{rotational boundary condition}), we arrive at the
symmetry-adapted tight-binding wavefunction
\begin{equation}
\arrowvert k,\mu \rangle = \frac{1}{\sqrt{N}} \sum_{r,\textbf{J}}
e^{i \ \phi_\textbf{J}(k,\mu)} \arrowvert \textbf{J},r \rangle
\label{TB symmetry adapted wavefunction}
\end{equation}
with the phase factor
$\phi_\textbf{J}(k,\mu) \equiv \hat{\textbf{k}}(k,\mu) \cdot \textbf{R}_{\textbf{J}}$ given by
\begin{eqnarray}
\nonumber
\phi_\textbf{J}(k,\mu) &=&
\frac{\pi \mu ((2n+m)j_1 + (2m+n)j_2))} { n^2 + nm + m^2 }
\\ &+&
\frac{\sqrt{3}\ a \ k}{2} \ \frac{m j_1 - n j_2}{\sqrt{n^2 +nm + m^2}}
\label{phase factor equation}
\end{eqnarray}

Substituting the symmetry-adapted tight-binding wavefunction
(\ref{TB symmetry adapted wavefunction}) into the Schr\"{o}dinger equation,
we obtain, for each value of $\mu$, a $2 \times 2$ matrix eigenvalue equation
for the electronic energies $E_{s \mu}(k)$, and the expansion coefficients
$C_r(s,\mu,k)$, namely
\begin{equation}
\sum_{r'} H_{r,r'} \ C_{r'} = E_{s \mu}(k) \sum_{r'} S_{r,r'} \ C_{r'}
\label{Electronic Matrix Eigenvalue Problem}
\end{equation}
where $s = v,c$ labels the valence
and conduction band states.  The $2 \times 2$ Hamiltonian and overlap matrices
are given by
\begin{equation}
H_{r,r'} = \sum_{\textbf{J}'} e^{i \phi_{\textbf{J}'}(k,\mu)}
\langle \textbf{0}, r \arrowvert H \arrowvert \textbf{J}', r' \rangle
\label{Electron Hamiltonian Matrix}
\end{equation}
and
\begin{equation}
S_{r,r'} = \sum_{\textbf{J}'} e^{i \phi_{\textbf{J}'}(k,\mu)}
\langle \textbf{0}, r \arrowvert \textbf{J}', r' \rangle
\label{Electron Overlap Matrix}
\end{equation}

In the sum over $\textbf{J}' = (j'_1,j'_2)$, in Eqs.~(\ref{Electron Hamiltonian Matrix})
and (\ref{Electron Overlap Matrix}) we keep only the on-site and third nearest neighbor
contributions since the parameterized matrix elements vanish
at distances beyond the third nearest neighbor distance in graphene.

Values of $(j_1,j_2)$ for the first to fourth nearest neighbors of the
A and B atoms in the two-atom unit cell $\textbf{J} = (0,0)$ are easy to work out and
can be found in Table 2 of Ref.~\onlinecite{Li04.657}.

\section{Phonon Modes}
\label{Appendix B}

Here we derive the block diagonal $6 \times 6$ dynamical matrices
and symmetry-adapted atomic displacement vectors for phonon
modes in a carbon nanotube.

The equilibrium position of the $r$-th atom in the $J$-th unit cell is
$\textbf{R}_{r\textbf{J}} = \textbf{R}_{\textbf{J}} + \tau_r$, where
$\tau_A = (- \frac{a}{2 \sqrt{3}}, 0)$ and $\tau_B = (\frac{a}{2 \sqrt{3}}, 0)$.
The position of the two-atom unit cell in unrolled graphene $xy$
coordinates, $\textbf{R}_{r\textbf{J}}$, is defined by Eq.~(\ref{graphene coordinates})
in Appendix \ref{Appendix A}.
The positions of the nearest neighbor atoms are $\textbf{R}_{r\textbf{J}}^{\alpha l}$,
where $\alpha$ denotes the nearest neighbor shell and $l$ labels the atoms within
the $\alpha$-th shell. The corresponding instantaneous atomic displacements are
$\textbf{U}_{r\textbf{J}}$ and $\textbf{U}_{r\textbf{J}}^{\alpha l}$.  In this
notation the bond stretching potential is given by \cite{Jiang06.235434}
\begin{equation}
\label{bond stretching potential energy}
V_s = \sum_{r\textbf{J} , \alpha l} \frac{\Gamma^\alpha_s}{2}
\Big(
\hat{\textbf{n}}^{\alpha l}_{r\textbf{J}} \cdot (\textbf{U}^{\alpha l}_{r\textbf{J}} -
\textbf{U}_{r\textbf{J}} )
\Big)^2
\end{equation}
where $\hat{\textbf{n}}^{\alpha l}_{r\textbf{J}}$ is a unit vector parallel to
$\textbf{R}_{r\textbf{J}}^{\alpha l} - \textbf{R}_{r\textbf{J}}$ and $\Gamma^\alpha_s$
are the bond stretching force constants.

The nearest neighbor in-plane bond bending potential is given by\cite{Lobo97.159}
\begin{eqnarray}
\nonumber
V_{bb}&=& \sum_{r\textbf{J}} {\sum_{l l'}}' \ \frac{\Gamma_{bb}}{2} \
\Big( \hat{\textbf{n}}^{1 l}_{r\textbf{J}} \cdot
(\textbf{U}^{1 l'}_{r\textbf{J}} - \textbf{U}_{r\textbf{J}})
\\ &&
+ \ \hat{\textbf{n}}^{1 l'}_{r\textbf{J}} \cdot
(\textbf{U}^{1 l}_{r\textbf{J}} - \textbf{U}_{r\textbf{J}})
\Big)^2
\label{bond bending potential in plane}
\end{eqnarray}
where $\Gamma_{bb}$ is the in-plane bond bending force constant. The prime on the
sum over nearest neighbors $l$ and $l'$ indicates that the $l = l'$ term is omitted.

The out-of-surface bond bending potential is given by\cite{Jiang06.235434}
\begin{equation}
\label{bond bending potential out of plane}
V_{rb} = \sum_{r\textbf{J}} \frac{\Gamma_{rb}}{2}
\Big(
\sum_{l} \hat{\varrho}_{r\textbf{J}} \cdot
( \textbf{U}^{1 l}_{r\textbf{J}} - \textbf{U}_{r\textbf{J}} )
\Big)^2
\end{equation}
where $\Gamma_{rb}$ is the out-of-plane bond bending force constant.  The unit
vector $\hat{\varrho}_{r\textbf{J}}$ is given by
\begin{equation}
\label{rho out of plane vector}
\hat{\varrho}_{r\textbf{J}} =
\frac{ \sum_l (\textbf{R}^{1 l}_{r\textbf{J}} - \hat{\textbf{z}} \
( \hat{\textbf{z}} \cdot \textbf{R}_{r\textbf{J}}) )}
{\arrowvert
\sum_l ( \textbf{R}^{1 l}_{r\textbf{J}} - \hat{\textbf{z}} \
( \hat{\textbf{z}} \cdot \textbf{R}_{r\textbf{J}}) )
\arrowvert}
\end{equation}
and for large diameter nanotubes is approximately equal to the radial outward
unit vector at $\textbf{R}_{r\textbf{J}}$. As pointed out in
Ref.~\onlinecite{Jiang06.235434} the form of $\hat{\varrho}_{r\textbf{J}}$
in Eq.~(\ref{rho out of plane vector}) is required to preserve rigid
rotational invariance.

The bond twisting potential is given by \cite{Jiang06.235434}
\begin{equation}
\label{bond twisting potential}
V_{tw}= \sum_{\langle i, j \rangle}
\frac{\Gamma_{tw}}{2}
\Big( \hat{\textbf{r}}_{\langle i, j \rangle} \cdot
((\textbf{U}_1 + \textbf{U}_2)-(\textbf{U}_3 + \textbf{U}_4))
\Big)^2
\end{equation}
where $\Gamma_{tw}$ is the bond twisting force constant. In the sum
$\langle i, j \rangle$ represents a bond between atom $i$ and one of its
nearest neighbors $j$ and $\hat{\textbf{r}}_{\langle i, j \rangle}$ is the
radially outward unit vector at the midpoint of the bond. The atomic displacements
$\textbf{U}_1 \ldots \textbf{U}_4$ are the displacements of the four atoms
attached to the $\langle i, j \rangle$ bond with $\textbf{U}_1$ related
to $\textbf{U}_2$ by a $C_2$ rotation about $\hat{\textbf{r}}_{\langle i, j \rangle}$.

The equations of motion for the carbon atoms are
\begin{equation}
\label{atomic equations of motion}
M \ \frac{\partial^2 \textbf{U}_{r\textbf{J}}}{\partial t^2} =
- \ \frac{\partial V}{\partial \textbf{U}_{r\textbf{J}}}
\end{equation}
where $M$ is the mass of a carbon atom and $V = V_s + V_{bb} + V_{rb} + V_{tw}$ is
the ionic vibrational potential.

Symmetry-adapted atomic displacement vectors, for each value of $\nu$, can be written as
\begin{equation}
\label{atomic displacement vectors}
\textbf{U}_{r\textbf{J}}= S(\theta_\textbf{J}) \ \hat{\textbf{e}}_r(q,\nu)
\ e^{ i (\phi_{\textbf{J}}(q,\nu) + \omega t) }
\end{equation}
where $q$ is the phonon wavevector, $\nu = 0 \ldots N_{\mbox{hex}}-1$ labels the
cutting lines, and $\omega$ is the phonon frequency. The unit vector $\hat{\textbf{e}}_r(q,\nu)$
is the phonon mode polarization for the $r$-th atom in the $\textbf{J} = (0,0)$
two-atom unit cell. The phase factor $\phi_{\textbf{J}}(q,\nu)$ is obtained
from Eq.~(\ref{phase factor equation}) by replacing $k,\mu$ with $q,\nu$.

The matrix $S(\theta_\textbf{J})$ is a unitary rotation matrix for rotations
about the nanotube axis (taken to be $z$), and it acts to rotate
$\hat{\textbf{e}}_r(q,\nu)$ to the $\textbf{J}$-th two-atom unit cell.
The counter-clockwise rotation angle is
$\theta_\textbf{J} = j_1 \ \theta_1 + j_2 \ \theta_2$, where
\begin{equation}
\label{theta1 equation}
\theta_1 = \frac{(2 n + m) \pi}{n^2 + n m + m^2}
\end{equation}
and
\begin{equation}
\label{theta2 equation}
\theta_2 = \frac{(2 m + n) \pi}{n^2 + n m + m^2}.
\end{equation}

Substituting the atomic displacement vectors (\ref{atomic displacement vectors}) into the
equations of motion (\ref{atomic equations of motion}), we obtain an eigenvalue
problem for the phonon frequencies and mode polarization vectors
\begin{equation}
\label{dynamical equations of motion}
\sum_{r'} \textbf{D}_{r,r'}(q,\nu) \ \hat{\textbf{e}}_r(\beta, q,\nu) =
M \omega_{\beta \nu}^2(q) \ \hat{\textbf{e}}_r(\beta,q,\nu)
\end{equation}
where $\textbf{D}_{r,r'}(q,\nu)$ is a $6 \times 6$ dynamical matrix for each
value of $\nu$ and $\omega_{\beta \nu}(q)$ is the phonon dispersion relation with
$\beta = 1 \ldots 6$.
Explicit expressions for the symmetry-adapted $6 \times 6$ dynamical matrix
in terms of the atomic force constants can be found in
Refs.~\onlinecite{Li04.657} and \onlinecite{Popov99.8355}.

The force constants obtained from the fit
shown in Fig.~\ref{Graphene phonon dispersion figure} are
$\Gamma^1_s = 237.6$, $\Gamma^2_s = 17.47$, $\Gamma^3_s = 2.895$, $\Gamma^4_s = 7.166$,
$\Gamma_{bb} = 18.37$, $\Gamma_{rb} = 16.67$ and $\Gamma_{tw}= 6.609$ in units
of Joules/meter$^2$.

\section{Electron-Phonon Interaction Matrix Elements}
\label{Appendix C}

The electron-phonon interaction
matrix element in Eq.~(\ref{second quantized Electron phonon Hamiltonian})
is given by
\begin{eqnarray}
\nonumber &&
M_{s, s', \beta}^{\mu,\nu}(k,q) =
- A_{\beta,\nu}(q) \sum_{r''} \hat{\textbf{e}}_{r''}(\beta, q, \nu)
\\ \nonumber &&
\cdot \sum_{r'\textbf{J}'} \
C_{r'}^{*}(s',\mu+\nu,k+q) \ e^{-i \phi_{\textbf{J}'}(k+q,\mu+\nu)}
\\ &&
\times \sum_{r\textbf{J}} \
C_{r}^{}(s, \mu, k) \ e^{i \phi_{\textbf{J}}(k,\mu)}
\ \vec{\lambda}(r'\textbf{J}', r'', r\textbf{J})
\label{Electron phonon matrix element}
\end{eqnarray}
where $C_{r}(s, \mu, k)$ are the electronic expansion coefficients in
Eq.~(\ref{Electronic Matrix Eigenvalue Problem}), $\phi_{\textbf{J}}(k, \mu)$ is
the phase function defined in Eq.~(\ref{phase factor equation}), and
$\hat{\textbf{e}}_{r}(\beta, q, \nu)$ are the phonon polarization vectors in
Eq.~(\ref{dynamical equations of motion}). In Eq.~(\ref{Electron phonon matrix element}),
we choose $\textbf{J}'' = \textbf{0}$ so that $S(\theta_{\textbf{J}''}) = 1$ and
restrict sums over $r\textbf{J}$ and $r'\textbf{J}'$ to fourth nearest neighbors
of $\textbf{R}_{r'',\textbf{0}}$.

The deformation potential vectors $\vec{\lambda}(r'\textbf{J}', r'', r\textbf{J})$
appearing in Eq.~(\ref{Electron phonon matrix element}) are three-center integrals
defined as
\begin{equation}
\vec{\lambda} = \int d\textbf{r} \
\varphi^{*}_{r'\textbf{J}'}(\textbf{r}-\textbf{R}_{r'\textbf{J}'}) \
\nabla v_c(\textbf{r}-\textbf{R}_{r''\textbf{0}}) \
\varphi_{r\textbf{J}}(\textbf{r}-\textbf{R}_{r\textbf{J}})
\label{deformation potential vector}
\end{equation}
where $\varphi_{r\textbf{J}}(\textbf{r}-\textbf{R}_{r\textbf{J}})$ is a $\pi$
orbital $\arrowvert r\textbf{J} \rangle$ localized at $\textbf{R}_{r\textbf{J}}$
and $v_c(\textbf{r}-\textbf{R}_{r''\textbf{0}})$ is a carbon atom potential
centered at $\textbf{R}_{r''\textbf{0}}$.

We evaluate the deformation potential vectors $\vec{\lambda}$ using the $2p_z$ atomic
wavefunctions and the screened atomic potential for carbon in Ref.~\onlinecite{Jiang05.205420}
obtained from an {\it ab initio} calculation in graphene.\cite{Gruneis2004}
The $\pi$ orbitals at $\textbf{R}_{r\textbf{J}}$ are
\begin{equation}
\varphi_{r\textbf{J}}(\textbf{r}) =
\left( \textbf{r} \cdot \hat{\rho}_{r\textbf{J}} \right)
\sum_l I_l \ \exp \left( -\frac{r^2}{2 \sigma_l^2} \right)
\label{pi orbital equation}
\end{equation}
where $\hat{\rho}_{r\textbf{J}}$ is a unit vector normal to the surface of the
nanotube at $\textbf{R}_{r\textbf{J}}$. Using Eq.~(\ref{pi orbital equation})
for the $\pi$ orbitals in the evaluation of $\vec{\lambda}$ allows us to take
the relative orientation of the $\pi$ orbitals on different sites into account.
Similarly the atomic potentials are taken to be
\begin{equation}
v_c(\textbf{r}) = \frac{1}{r} \sum_l v_l \ \exp \left( -\frac{r^2}{2 \tau_l^2} \right)
\label{atomic potential equation}
\end{equation}
Values of $I_l$, $v_l$, $\sigma_l$ and $\tau_l$ are tabulated in Table I of
Ref.~\onlinecite{Jiang05.205420}. Substituting the expansions (\ref{pi orbital equation})
and (\ref{atomic potential equation}) into Eq.~(\ref{deformation potential vector})
we obtain an expansion for $\vec{\lambda}$ in terms of three-dimensional integrals. In the case of
graphene, the three-dimensional integrals can be done analytically.\cite{Jiang05.205420,Gruneis2004}
For the nanotube case we can reduce the three-dimensional integrals to one-dimensional
integrals that can be done numerically.


\begin{acknowledgments}
This work was supported by the National Science Foundation under grant numbers
DMR-0325474, OISE-0530220, DMR-0706313, and the
Robert A.~Welch foundation under grant number C-1509.
Y. S. Lim and K. J. Yee are supported by a Korea Science and
Engineering Foundation (KOSEF) grant funded by the Korean Government (Most)
(R01-2007-000-20651-0). R. Saito is supported by the Ministry of Education,
Culture, Sports, Science and Technology-Japan (MEXT) under grant
numbers 16076201 and 20241023.
\end{acknowledgments}

\bibliography{paper}

\end{document}